\documentclass[aps,pre,twocolumn,showpacs,floatfix]{revtex4}
\usepackage{graphicx,color}
\usepackage{amsmath,amssymb}

\def\beq{\begin{equation}}
\def\eeq{\end{equation}}
\def\beqa{\begin{eqnarray}}
\def\eeqa{\end{eqnarray}}

\def\bdi{\begin{displaymath}}
\def\edi{\end{displaymath}}

\newcommand{\figphasepure}
{\begin{figure}[htbp] 
\centering
\includegraphics[width=3.2in]{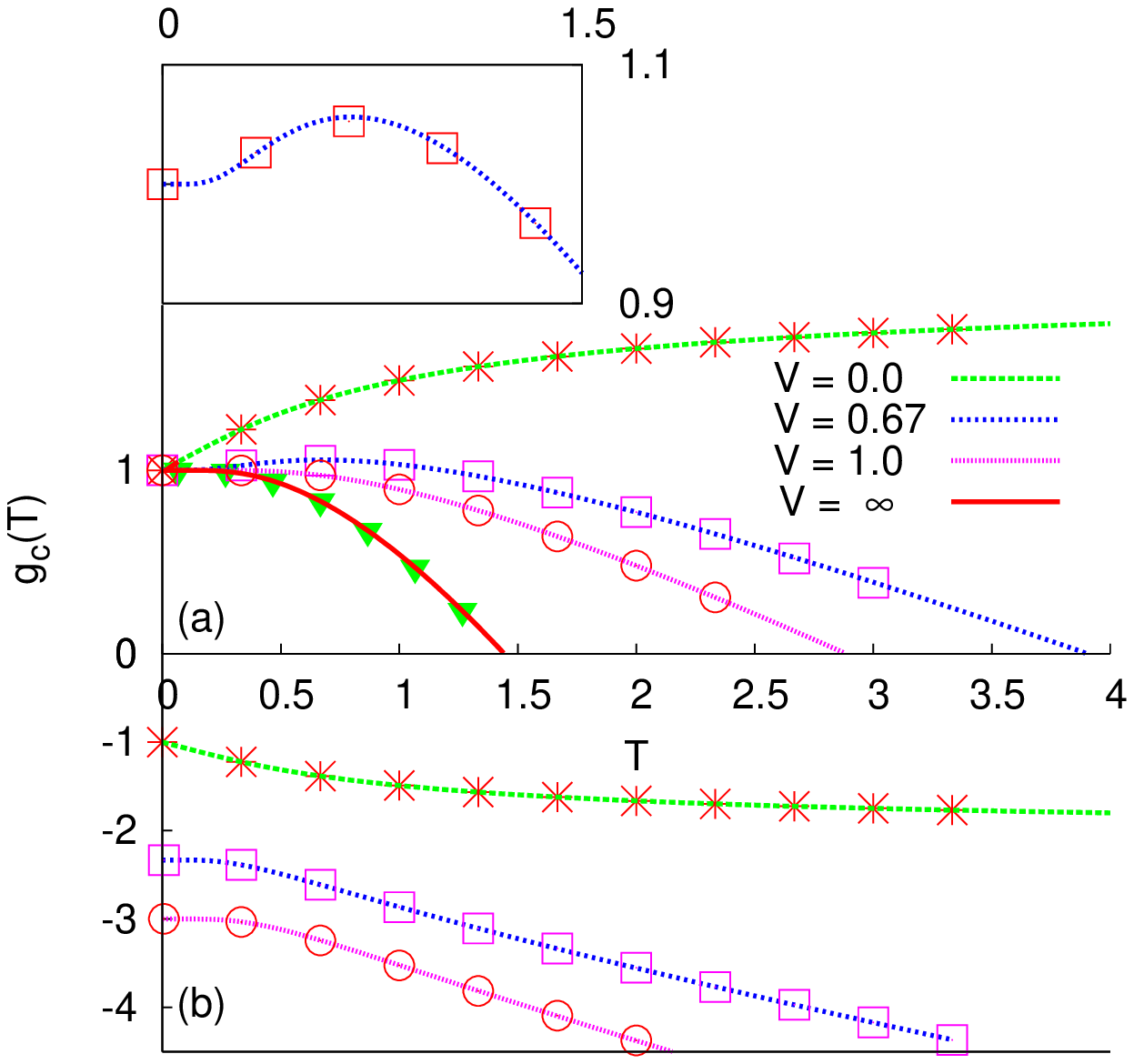} 

\caption{ Dimensionless force ($g$) versus temperature ($T$) phase diagram for an adsorbed polymer.  Each lattice site on one side ($x<0$) of the wall has a repulsive potential of strength $V >0$. Positive $g$ corresponds to case of pulling the polymer away from the wall on favorable side while the negative $g$ (note the different scale) 
is for pulling on the wrong side. The two limiting cases are $V \rightarrow 0$ (softwall) and $V \rightarrow \infty$ (hardwall). The points are from numerics and the lines are the exact analytical results. Note the reentrance on the positive side for $V=0.67$ (shown in inset). }

\label{fig:phasepure}
\end{figure}
}

\newcommand{\purecoll}
{\begin{figure}[htbp]
\includegraphics[width=3.2in]{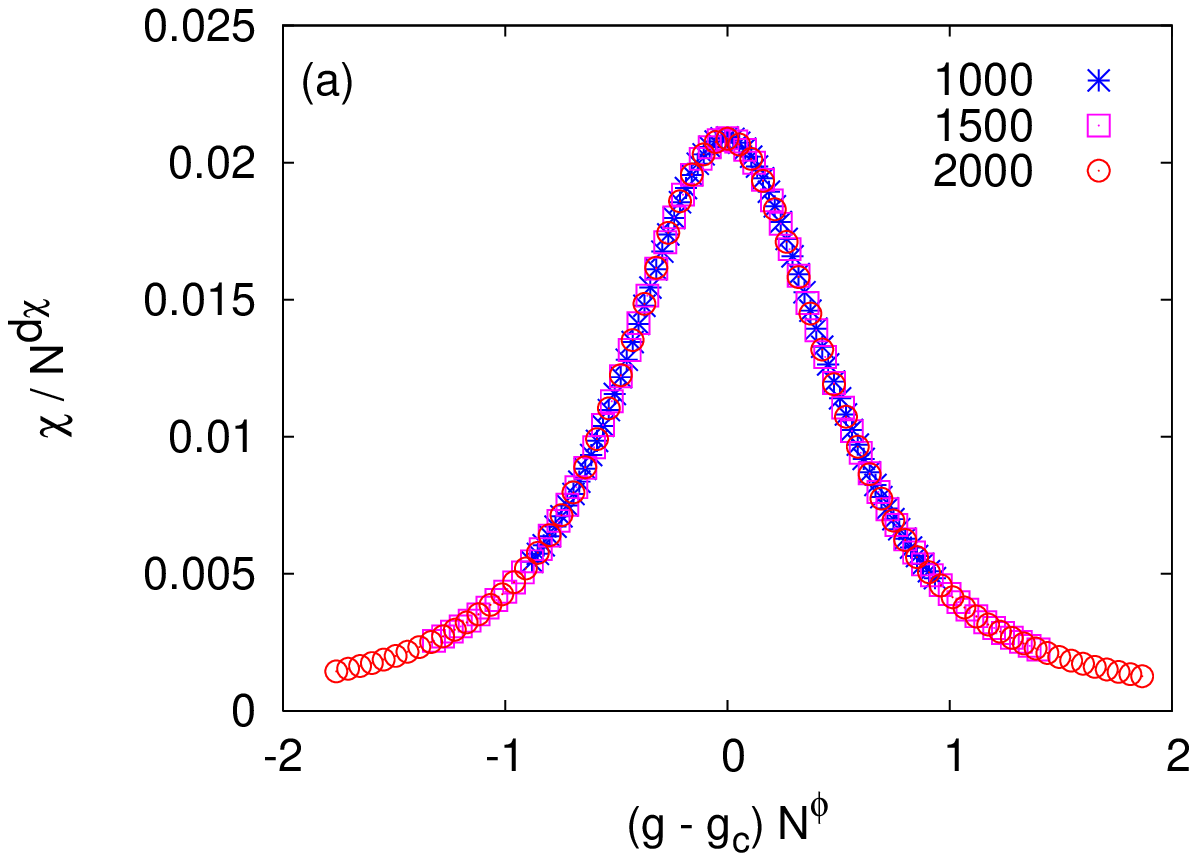}
\includegraphics[width=3.2in]{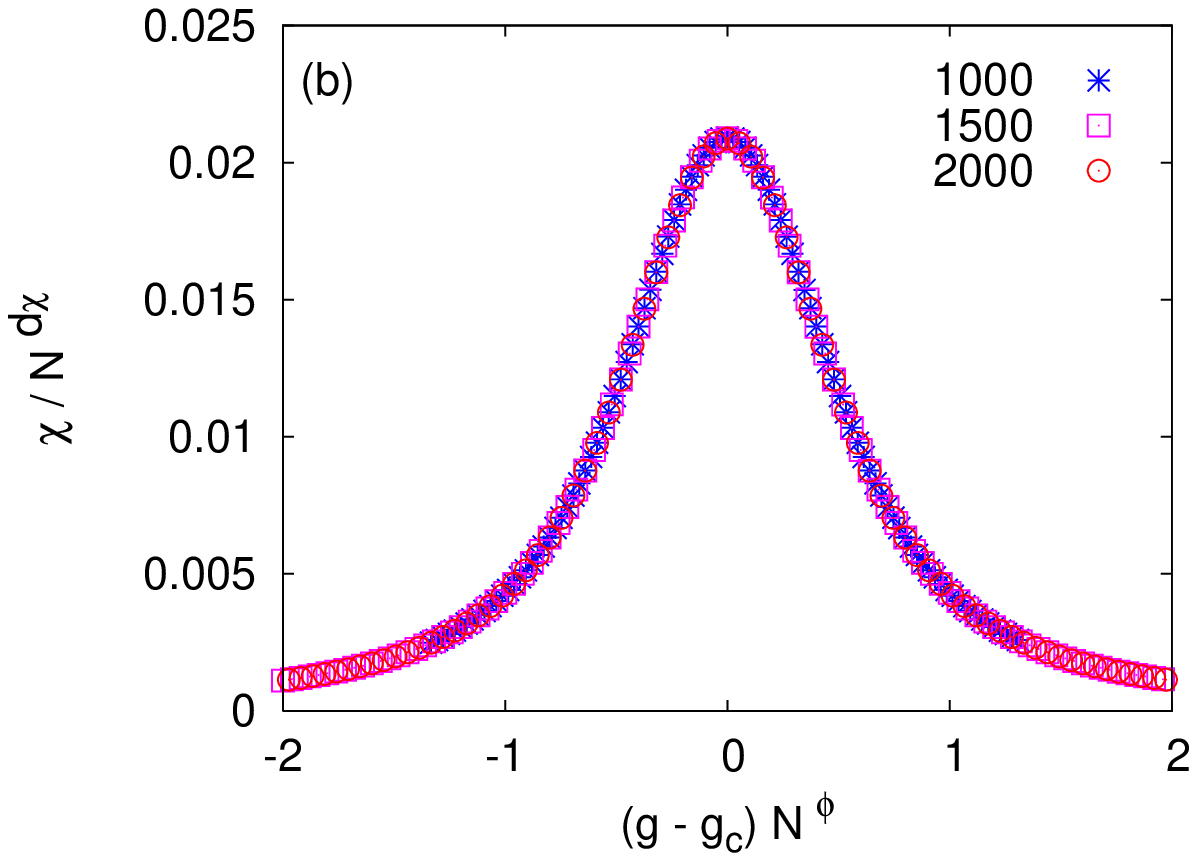} 
\caption{ Data collapse of extensibility at $\beta \epsilon =15$ ($T=0.067$) for the
  pure case (a) softwall (b) hardwall. The lengths of the chain used are $N=1000, 1500, 2000$. For both cases, the exponents are $d_{\chi}=2$ and $\phi=1$}
\label{fig:purecoll}
\end{figure}
}

\newcommand{\isotherm}
{\begin{figure}[htbp] 
\includegraphics[width=3.2in]{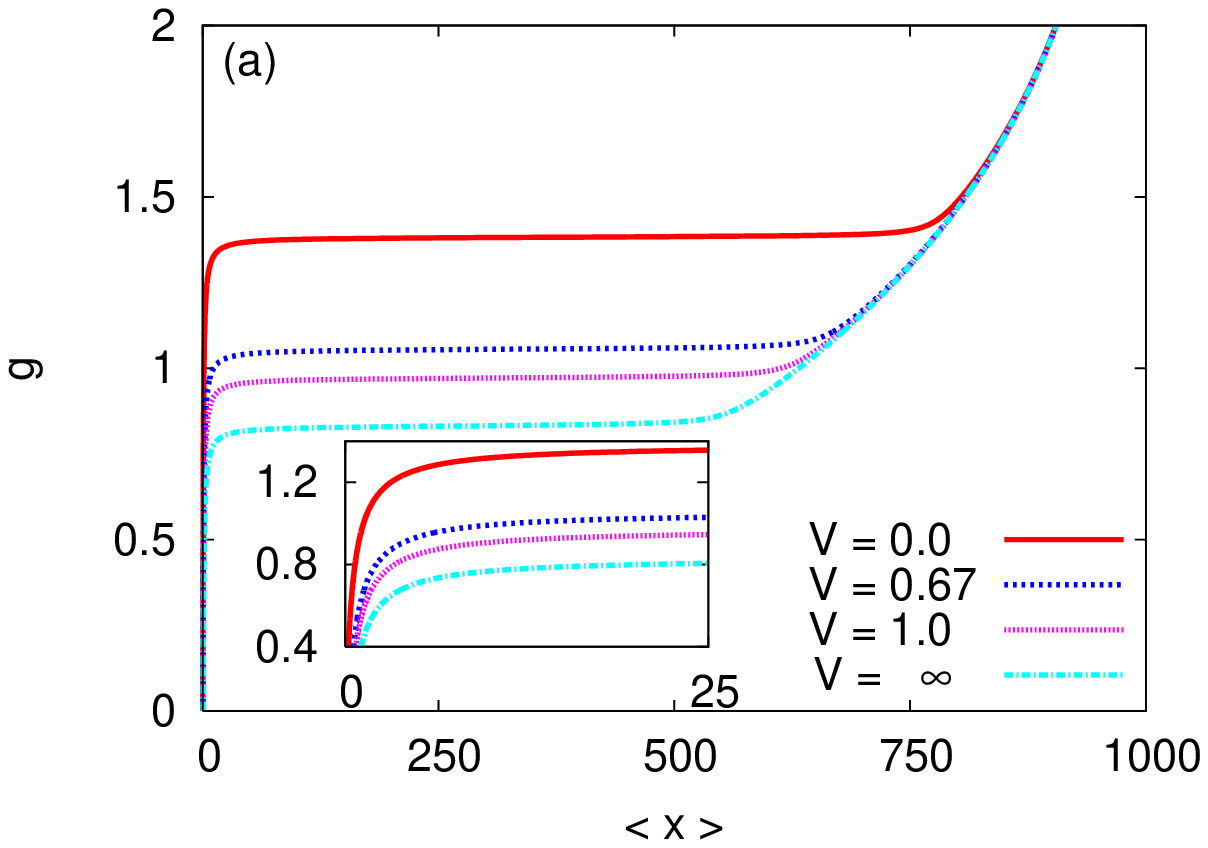}
\includegraphics[width=3.2in]{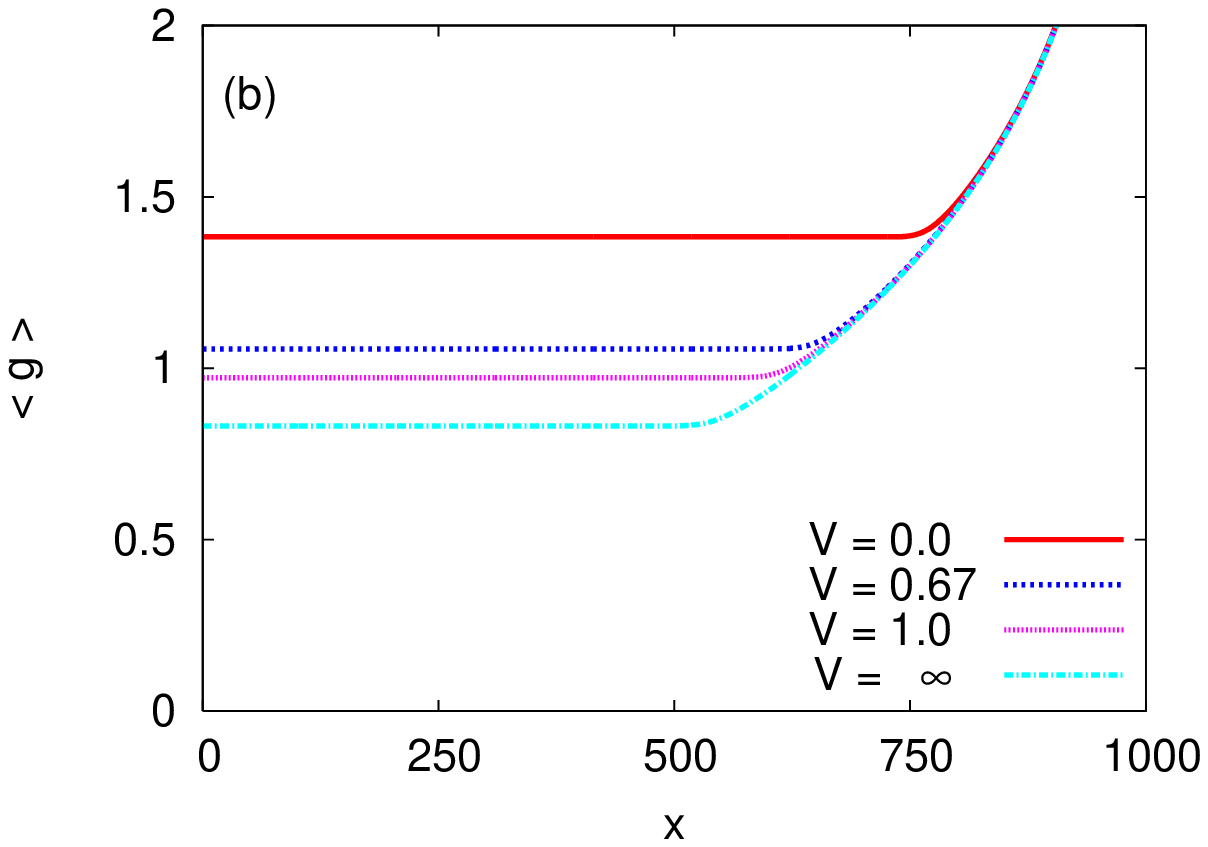} 

\caption{ Force distance isotherms in (a) fixed force ensemble (b) fixed distance ensemble, for a polymer of length $N = 2000$ at  $\beta \epsilon = 1.5$ ($T=0.067$).
  Each lattice site on one side of the wall contains a repulsive potential of strength ``V '' and polymer is pulled away from the  wall on other side.  The inset in (a) shows the knees which developed in fixed force ensemble for low $x$ but absent in 
  fixed distance ensemble.}
\label{fig:isotherm}
\end{figure}
}

\newcommand{\figspheat}{\begin{figure}[htbp]
\includegraphics[width=3.2in]{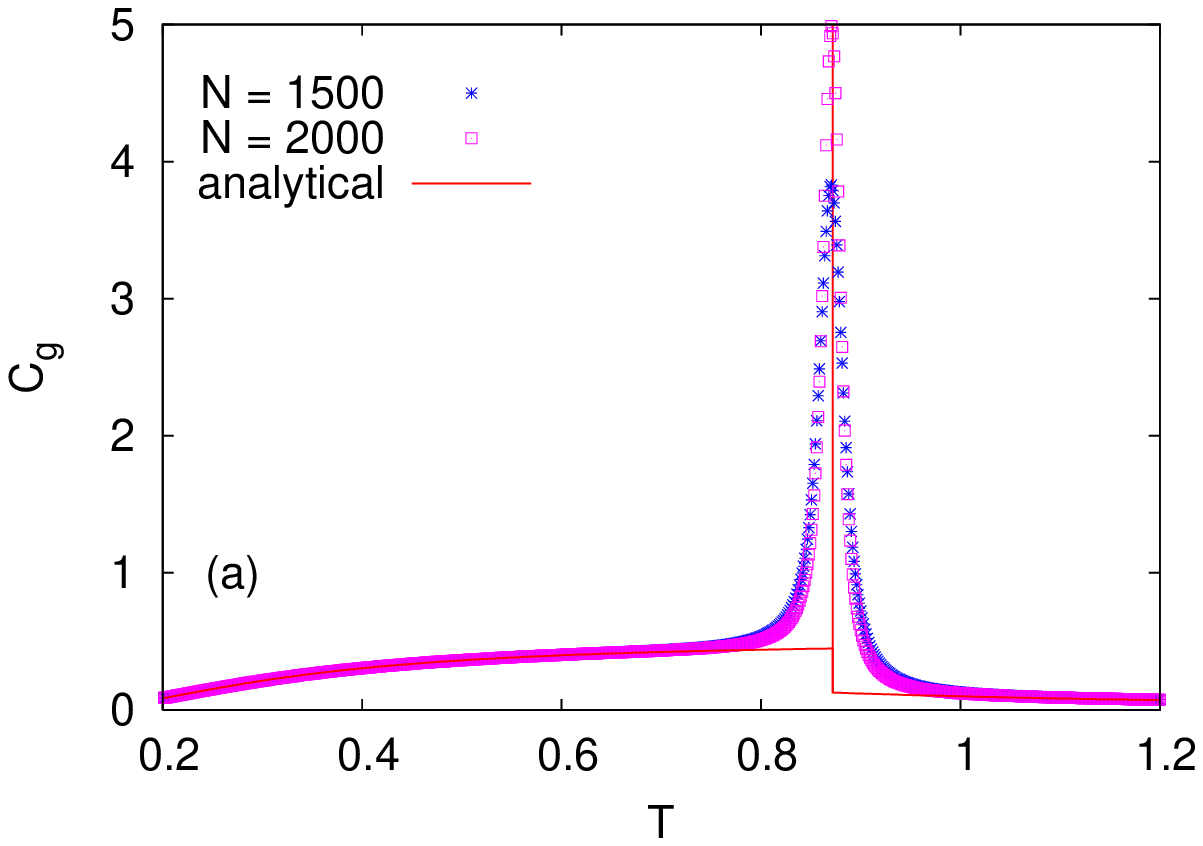}
\includegraphics[width=3.2in]{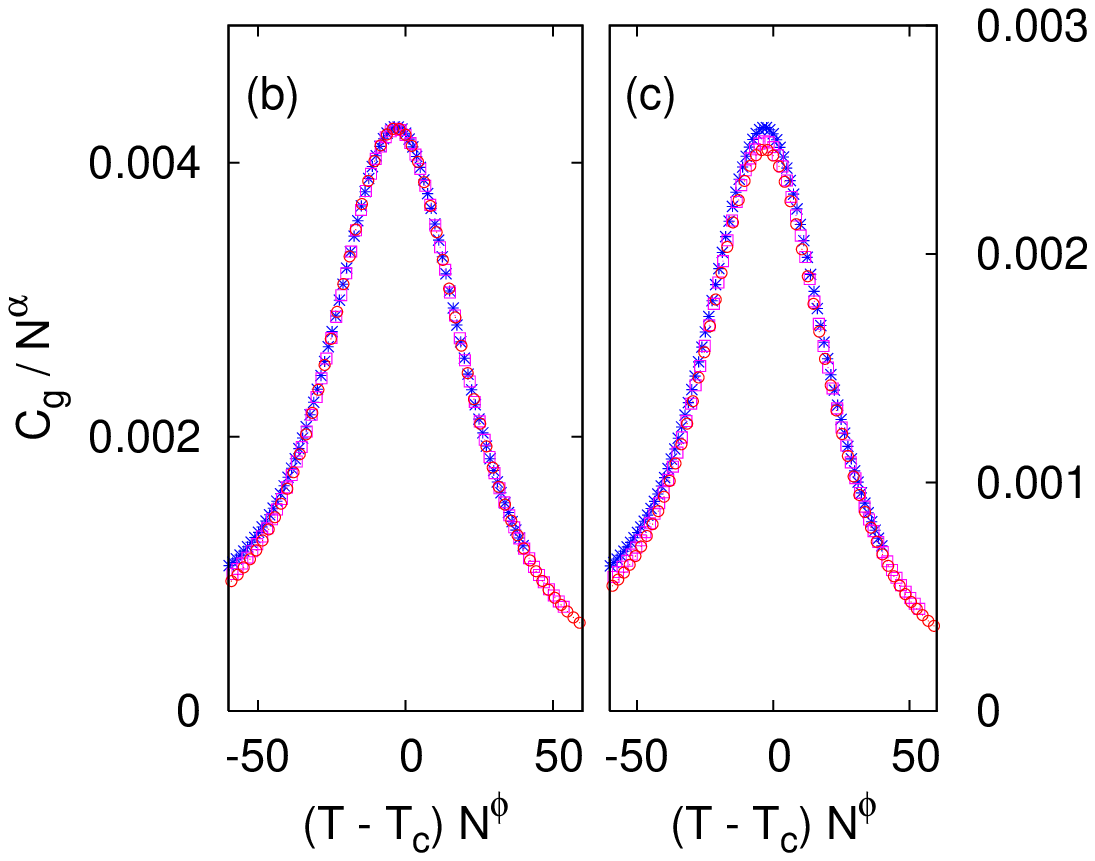}
\caption{ Specific heat up to a multiplicative constant for $g = 1.0$.  (a) There is a peak developing  as $N$ increases, over and above a discontinuity at $T_c(g) = 0.87204$.  (b, c) Data  collapse at $T_c(g)$ for chains of length $N = 1500, 2000, 2500$,  with $\phi_t=1$ and $\alpha\phi_t=0.93$ in (b) but $\alpha\phi_t=1$ in (c). }
\label{fig:spheat}
\end{figure}
}

\newcommand{\figa}{\begin{figure}[htbp]
\includegraphics[width=3.2in]{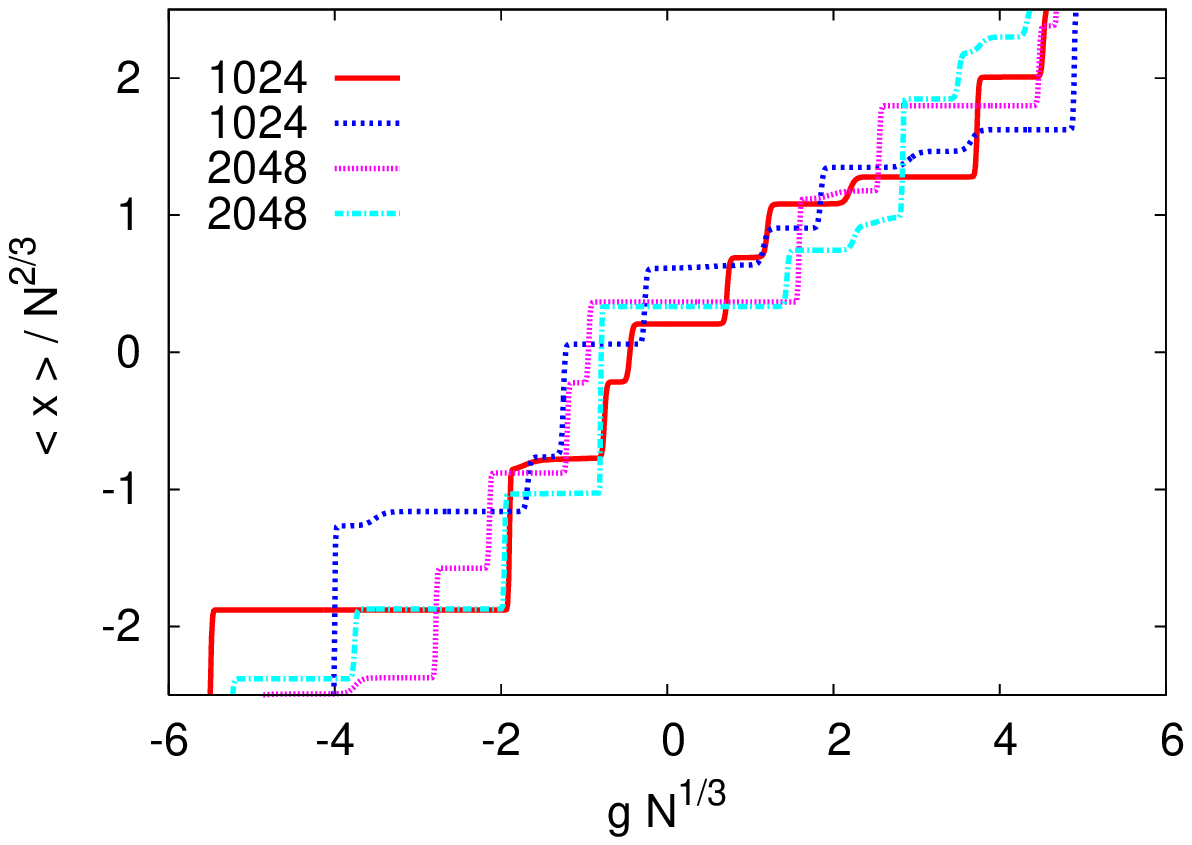}
\includegraphics[width=3.2in]{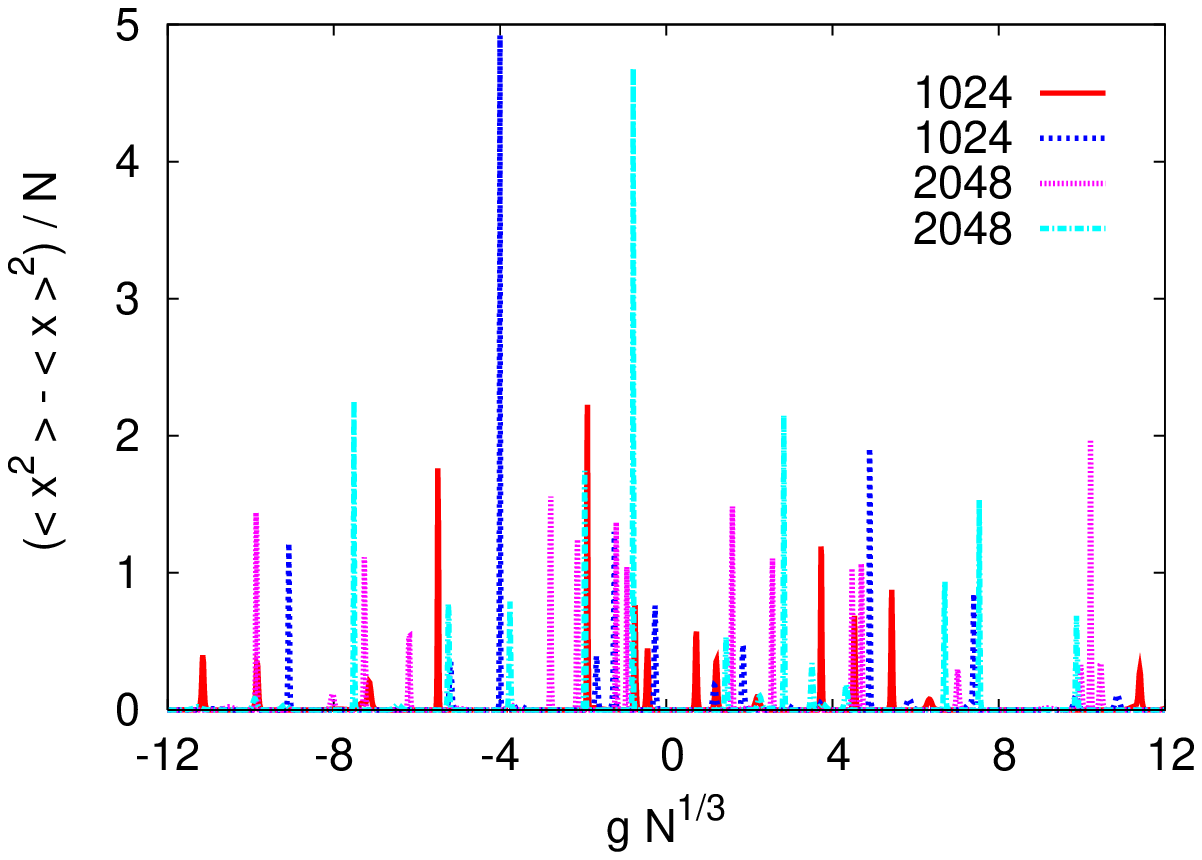}
\caption{ (a) Re-scaled force $g N^{1/3}$ versus re-scaled average distance $\langle x \rangle /N^{2/3}$, from the wall for four different samples of size $N=1024$ and $N=2048$ at  $\beta \epsilon =15$ ($T=0.067$) for the disorder strength $\Delta = 2$ and there is no binding to the wall ($\epsilon=0$). (b) Extensibility for the same samples.}
\label{fig:a}
\end{figure}
}

\newcommand{\figb}{\begin{figure}[htbp]
\includegraphics[width=3.2in]{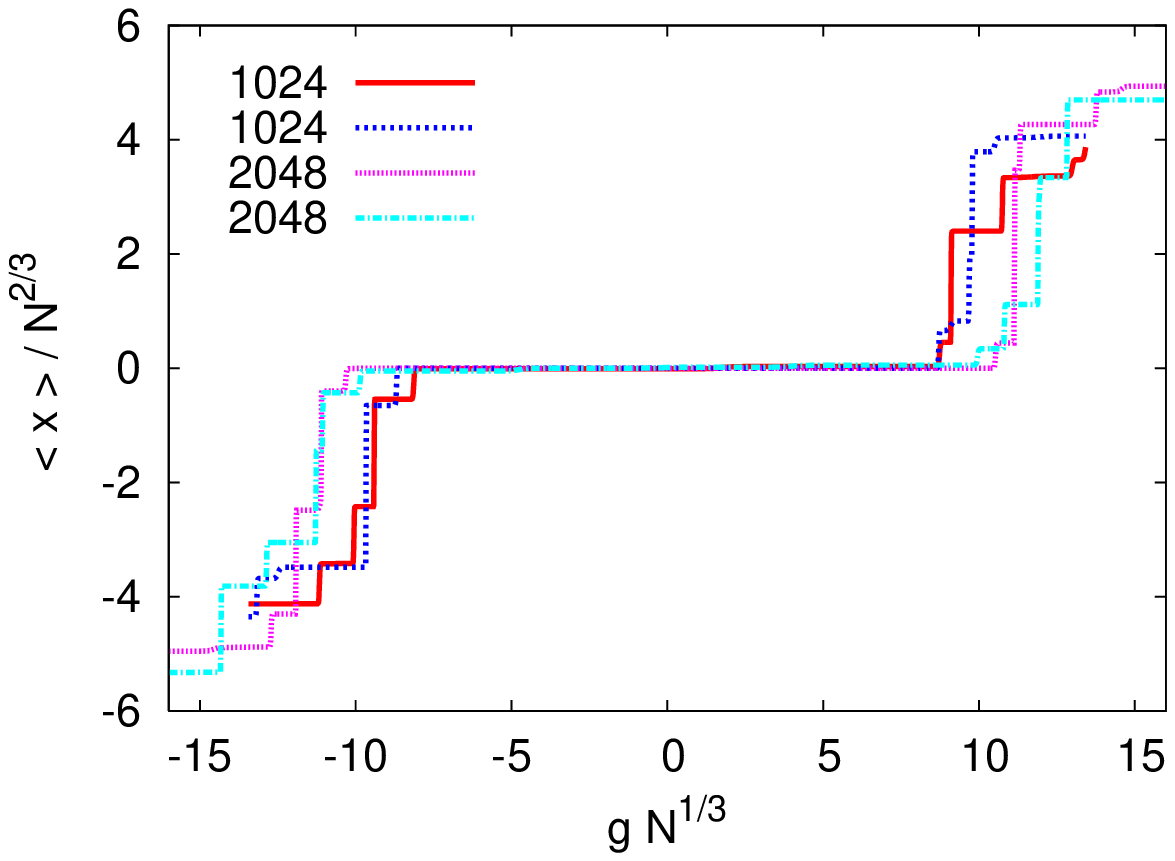}
\includegraphics[width=3.2in]{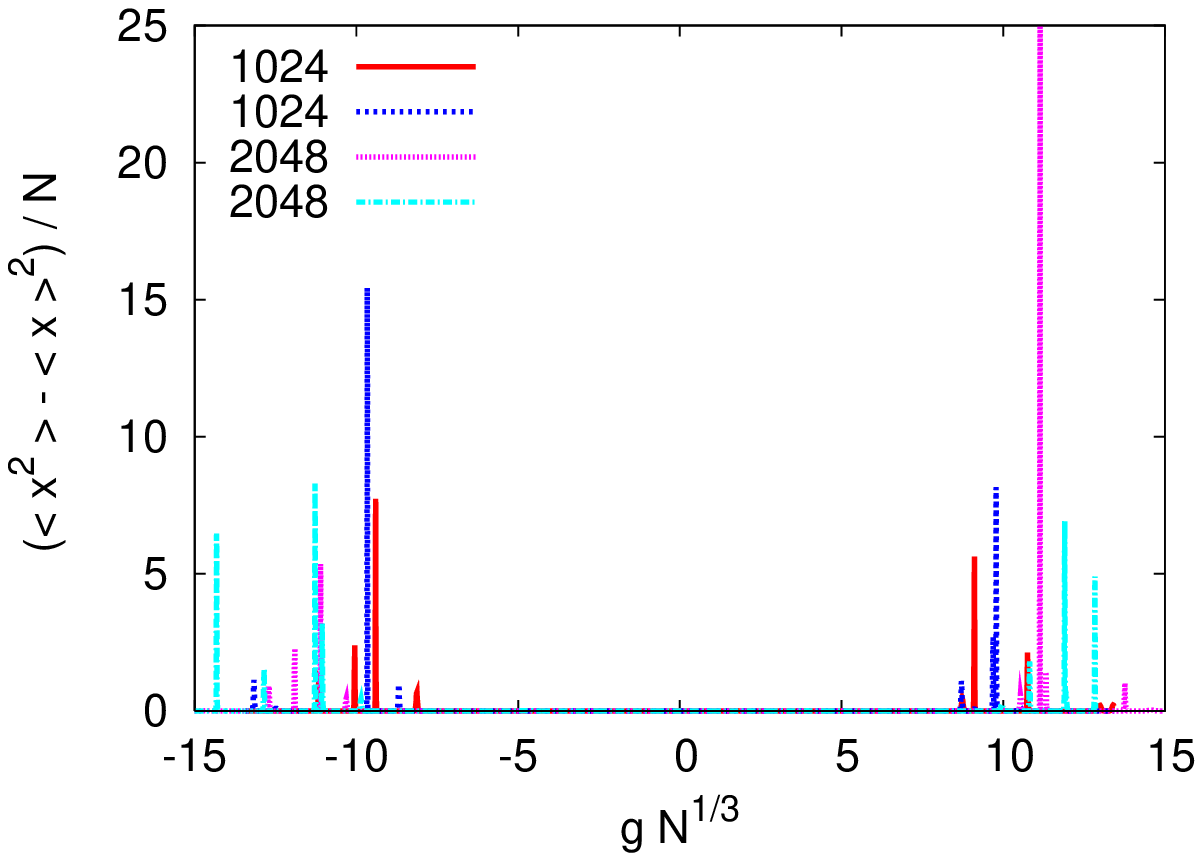}
\caption{ (a) Re-scaled force $g N^{1/3}$ versus re-scaled average distance $\langle x \rangle /N^{2/3}$, from the wall for four different samples of size $N=1024$ and $N=2048$ at $\beta \epsilon = 15$ ($T=0.067$) for the disorder strength $\Delta = 2$ 
when there is binding to the wall (b) Extensibility for the same samples.}
\label{fig:b}
\end{figure}
}

\newcommand{\figc}{\begin{figure}[htbp]
\includegraphics[width=3.2in]{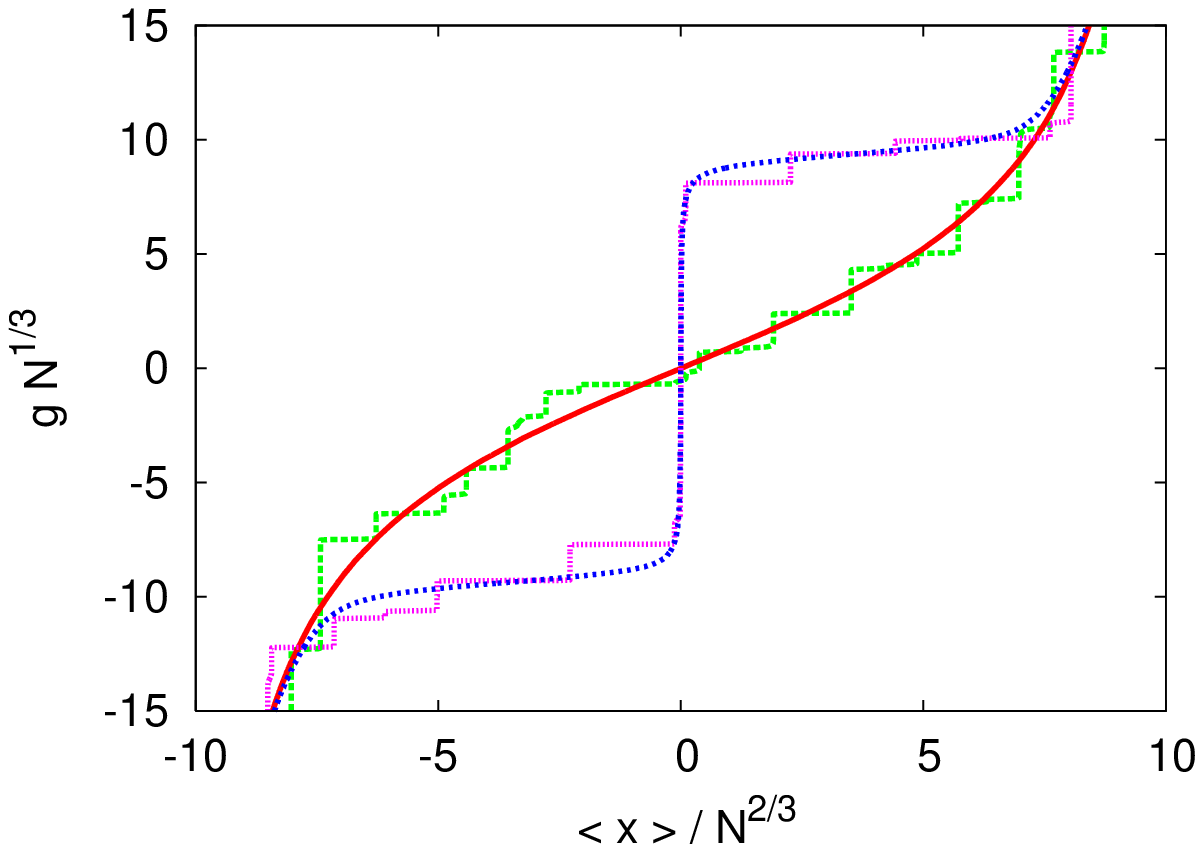}
\caption{ (a) Re-scaled force $g N^{1/3}$ versus re-scaled average distance  $\langle x \rangle /N^{2/3}$, from the wall for a polymer of length   $N=2048$ at $\beta \epsilon
 = 15$ ($T = 0.067$). The disorder is kept same (at a disorder strength   $\Delta =2$) for both binding and no binding ($\epsilon=0$)  to the wall. The steps are for a single sample and smooth curves are  after averaging over $10^4$ samples.}
\label{fig:c}
\end{figure}
}

\newcommand{\figprobsteps}{\begin{figure}[htbp]
\includegraphics[width=3.2in]{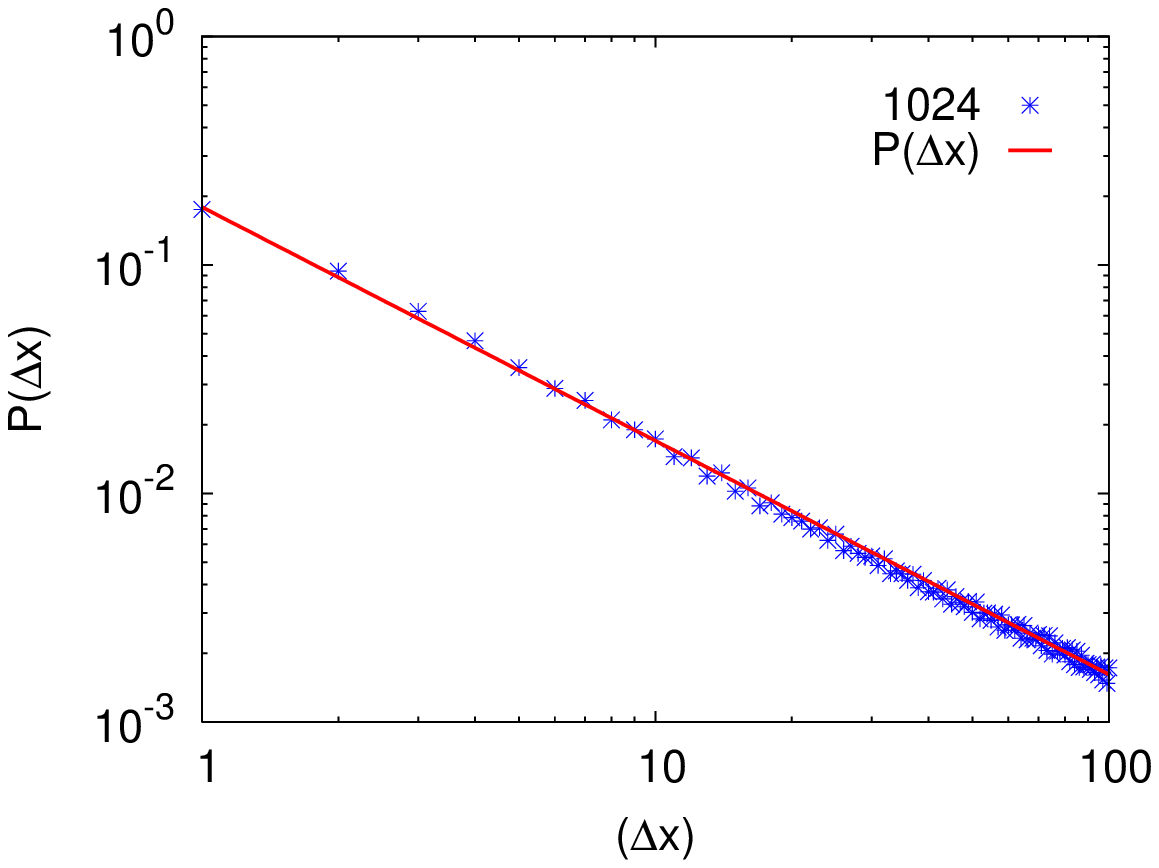}
\caption{Log log plot of the probablity distribution $P({\langle \Delta x \rangle})$ 
  of the rescaled average step height $\langle \Delta x\rangle$.  This  is for a polymer of length $N=1024$ in a fixed force ensemble at $\beta \epsilon = 15$  ($T=0.067$),
      disorder strength $\Delta=2$ and there is no binding to the wall ($\epsilon=0$). $10^5$ samples used to generate the  statistics. The slope of the fitted line is $1$. } 
\label{fig:d}
\end{figure}
}

\newcommand{\figpseudogc}{\begin{figure}[htbp]
\includegraphics[width=3.2in]{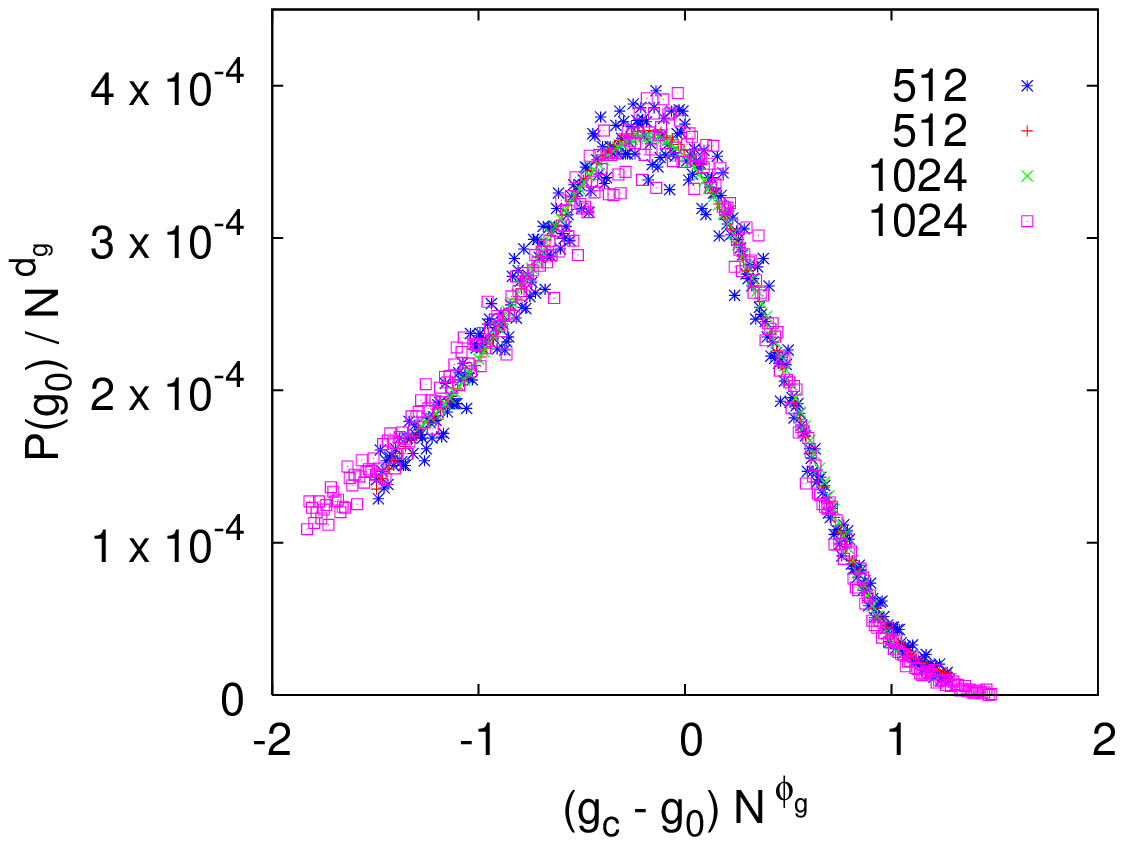}
\caption{Data collapse for the probability distribution of pseudo critical force
$P(g_0)/N^{d_g}$ versus $(g_c-g_0)N^{\phi_{g}}$ for the hardwall case for
$N=512, 1024$.  $d_g=0.41$ and $\phi_g= 0.31$. The averaging is done over $10^5$ samples.}
\label{fig:gc}
\end{figure}
}

\newcommand{\figcorr}{\begin{figure}[htbp]
\includegraphics[width=1.65in]{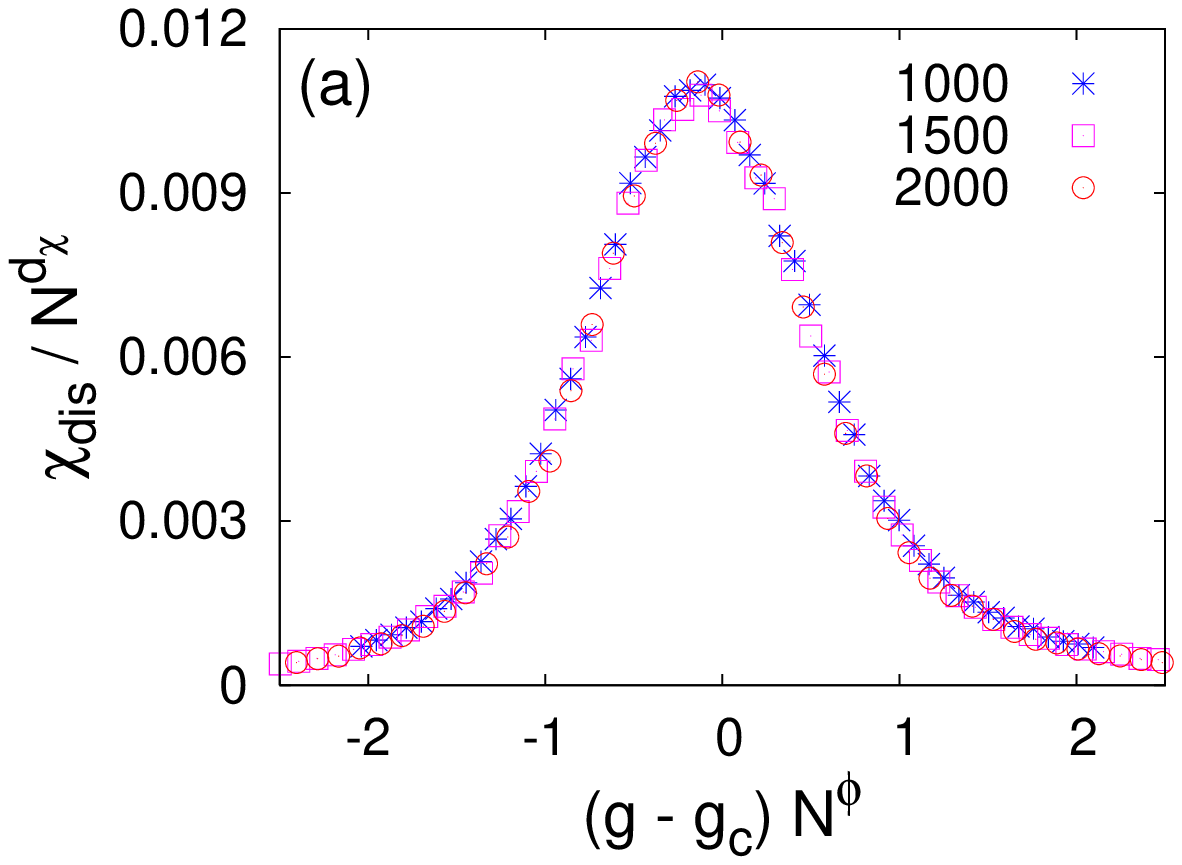}
\includegraphics[width=1.65in]{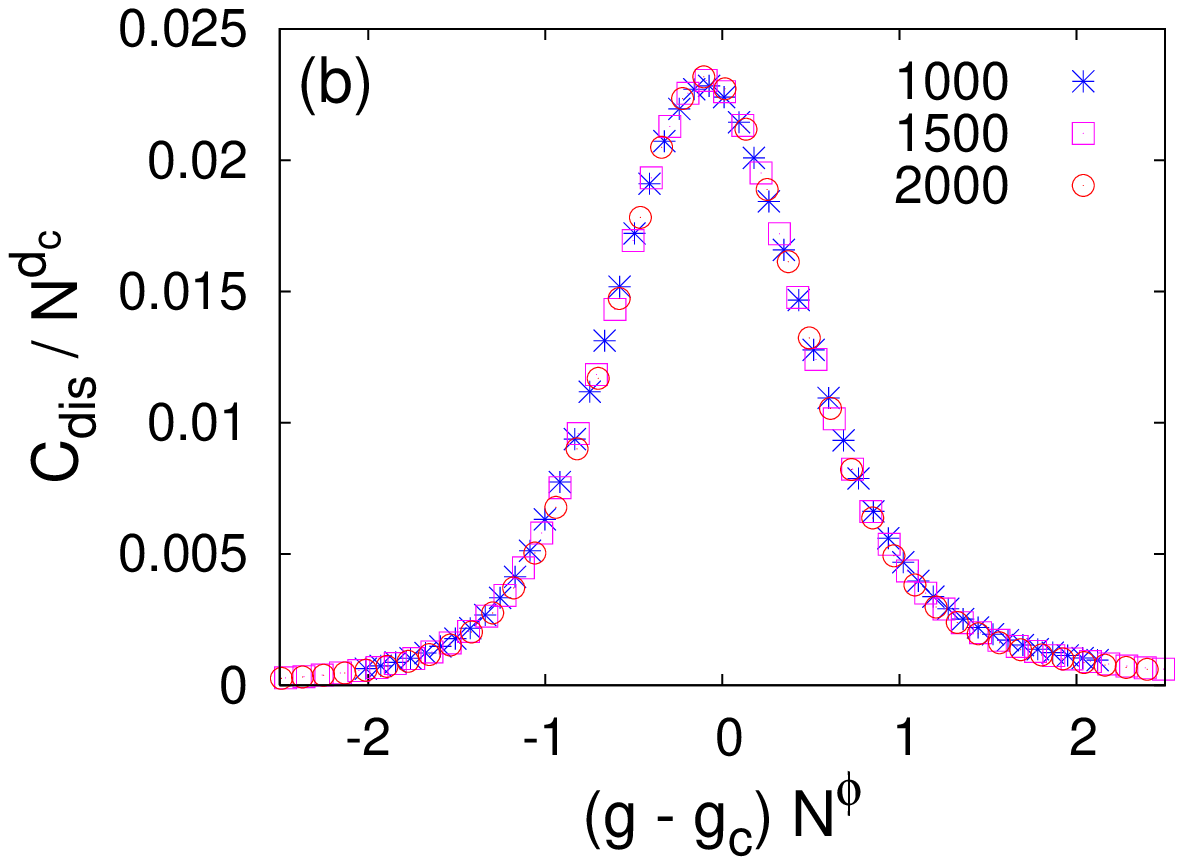}
\includegraphics[width=1.65in]{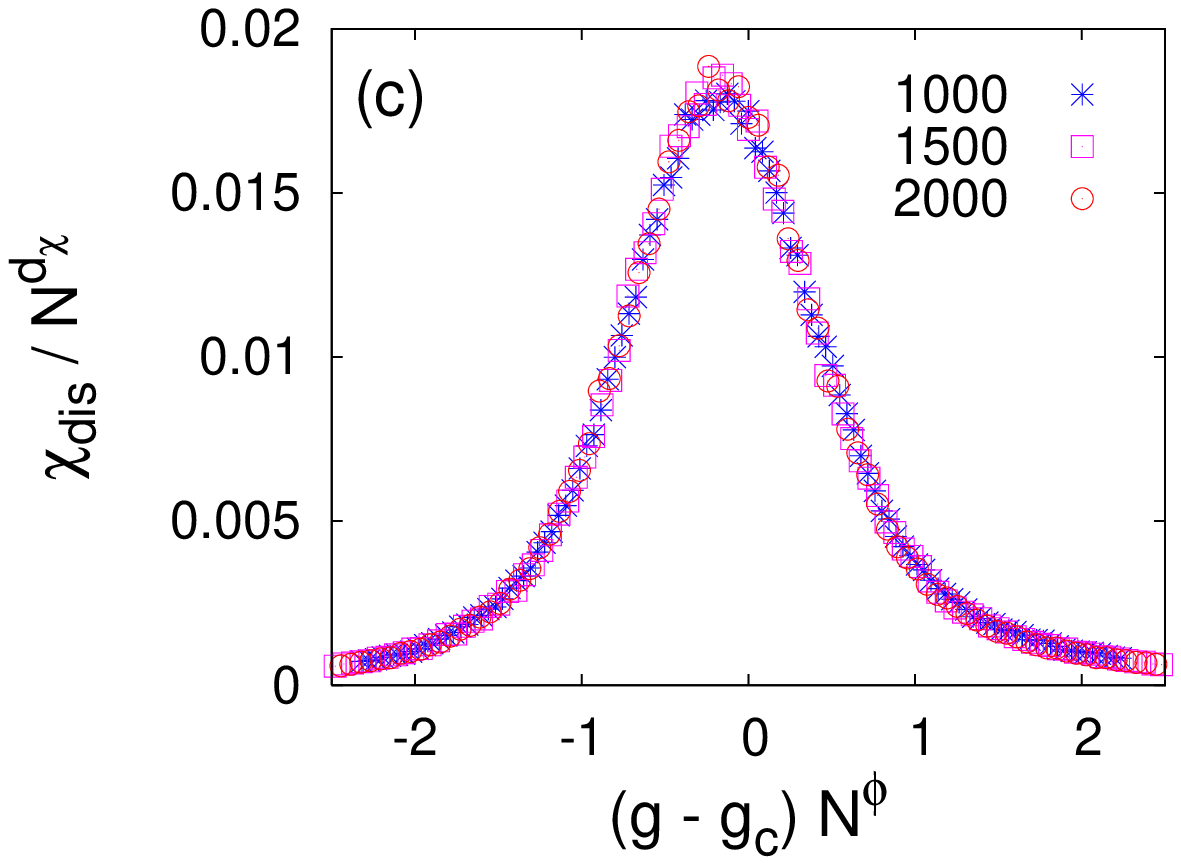}
\includegraphics[width=1.65in]{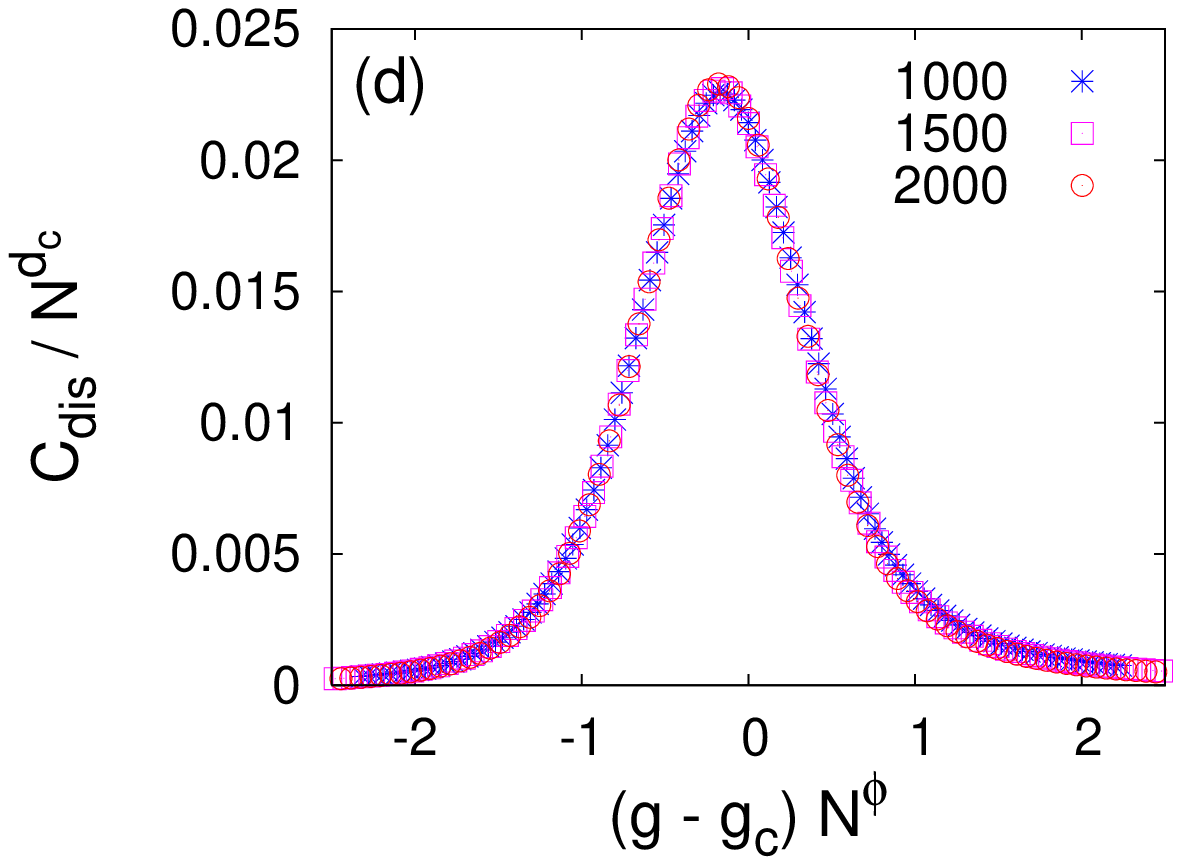}
\caption{ Data collapse for two different correlation functions for the random system. (a), (b) are for hardwall and (c), (d) are for softwall. The chain length $N=1000, 1500$ and $2000$. In (a) and (c) the disorder averaging is done at the response function level of an individual sample (by definition it is the extensibility upto a factor of
  $T^{-1}$). The exponents are $d_{\chi} =1.5$, $\phi = 0.5$. In (b) and (d) averaging is done at moment level. The exponents are $d_{\cal C} =2.0$, $\phi = 0.5$. For all these correlation functions, the disorder strength  $\Delta =2/3$ and averaging is
  done over $10^5$   realizations at $\beta\epsilon =15$ ($T = 0.067$).}
\label{fig:coll}
\end{figure}
}

\newcommand{\figfordis}{\begin{figure}[htbp]
\includegraphics[width=3.2in]{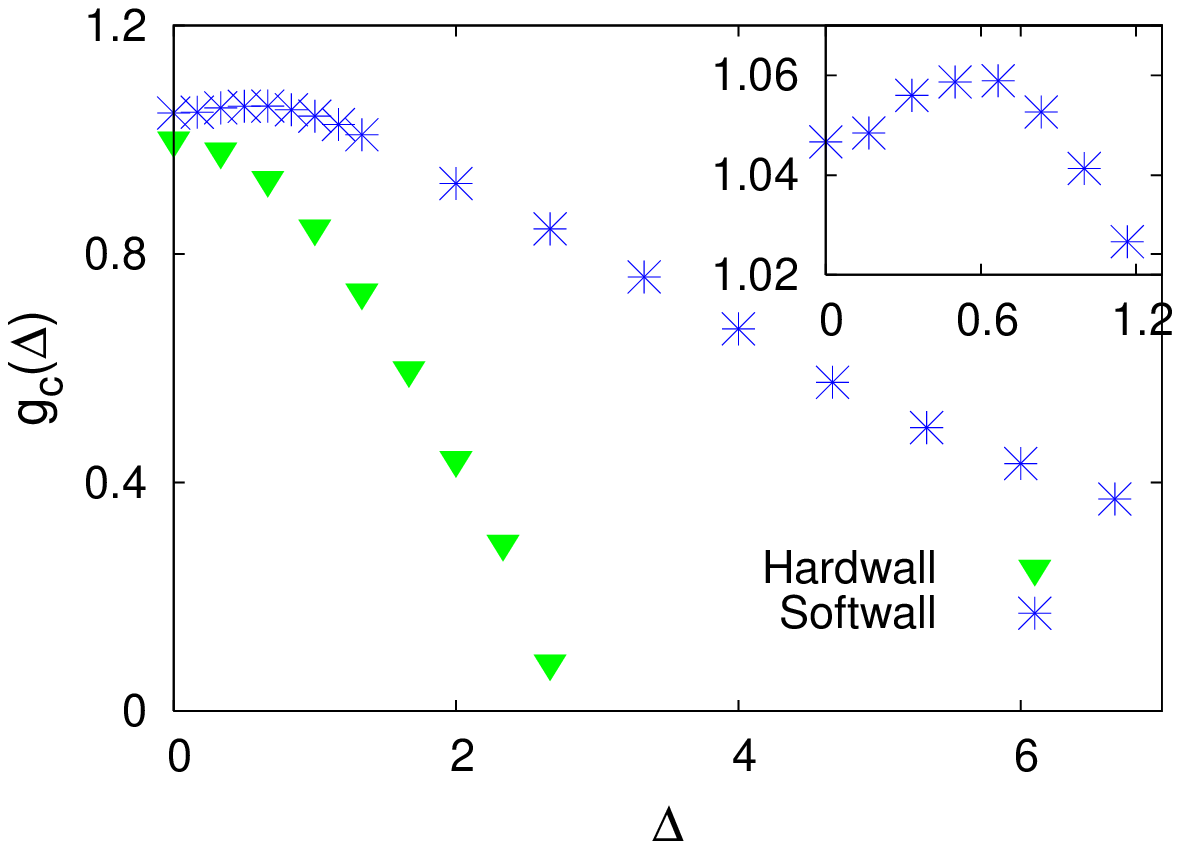}
\caption{ Critical force vs disorder strength  for both the hardwall and the softwall cases at  $\beta \epsilon = 15$, i.e. ($T=0.067$). The thin slice of reentrance is shown in the  inset.} 
\label{fig:fd}
\end{figure}
}

\newcommand{\figthree}{\begin{figure}[htbp]
\includegraphics[width=3.2in]{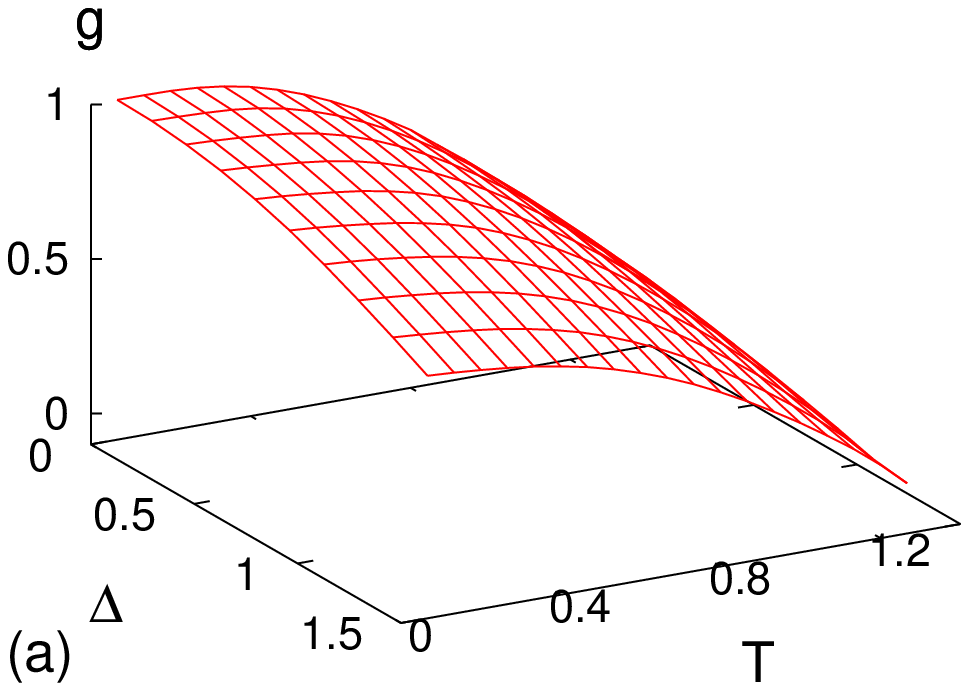}
\includegraphics[width=3.2in]{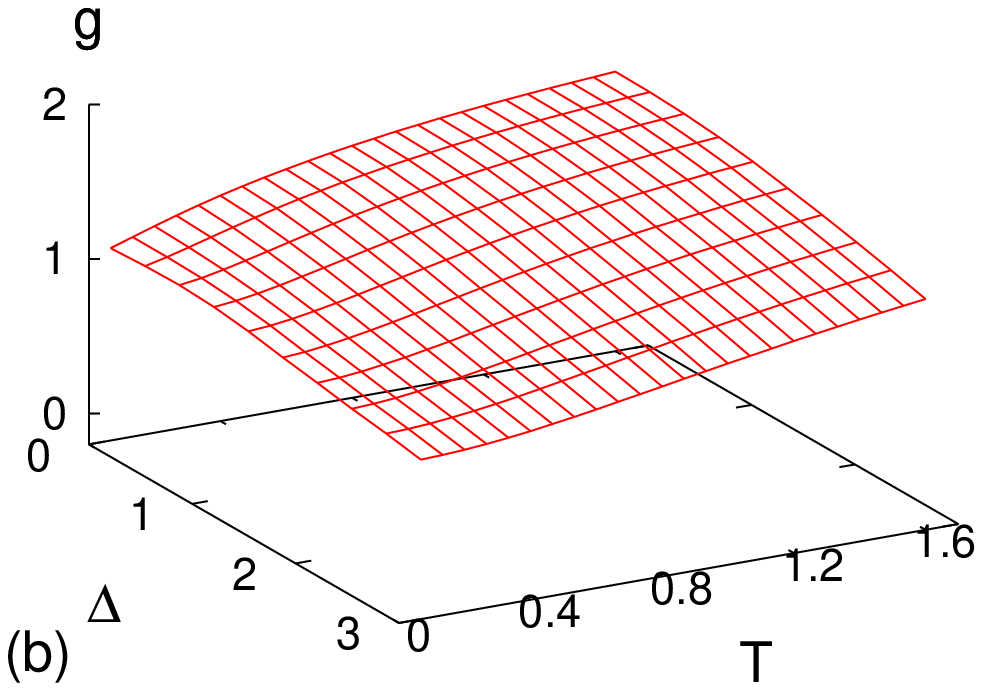}
\caption{ Phase surface for (a) hardwall and (b) softwall in the presence of disorder.}
\label{fig:3}
\end{figure}
}

\newcommand{\figmodel}{\begin{figure}[htbp]
\includegraphics[width=1.65in]{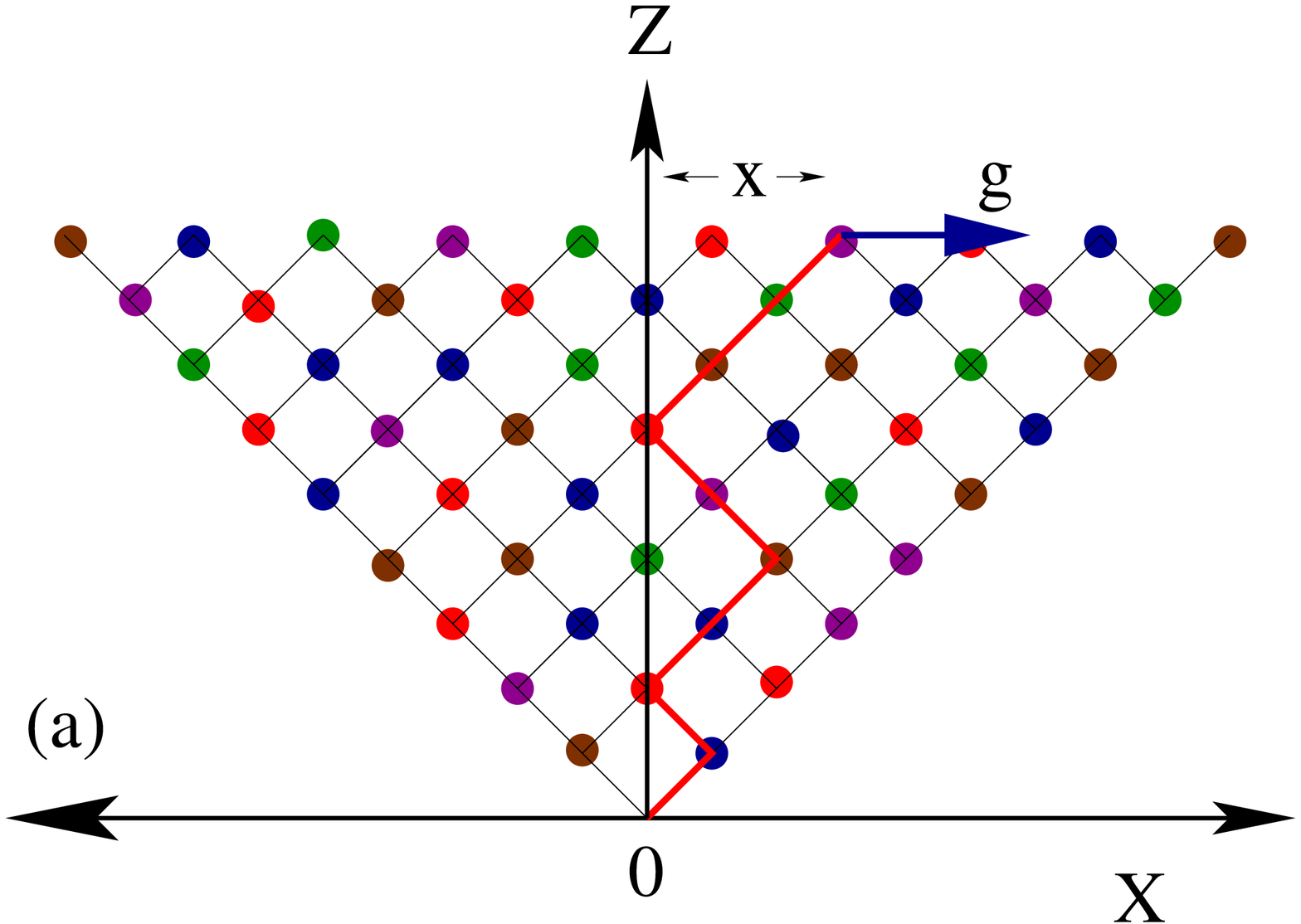}
\includegraphics[width=1.65in]{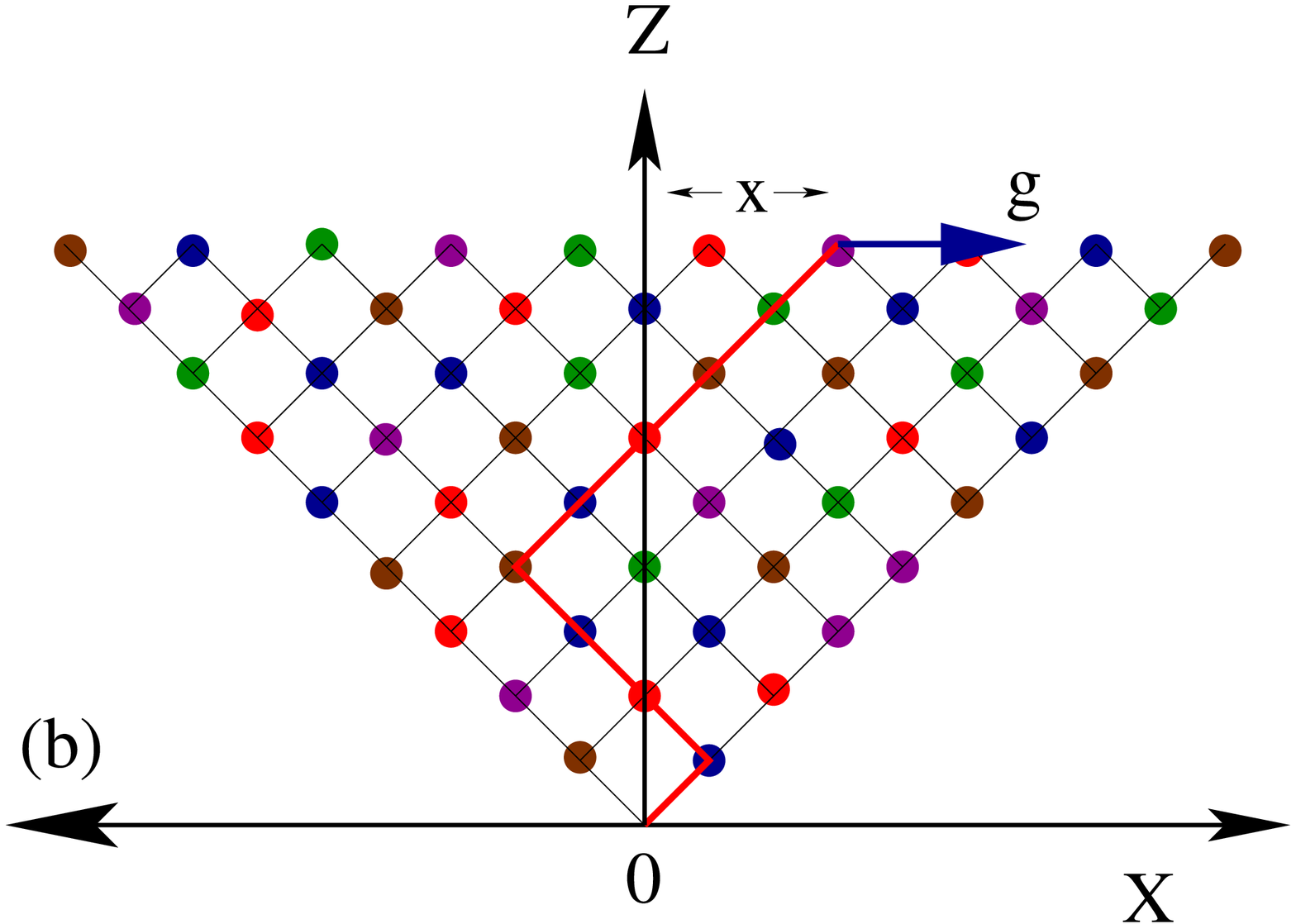}
\includegraphics[width=1.65in]{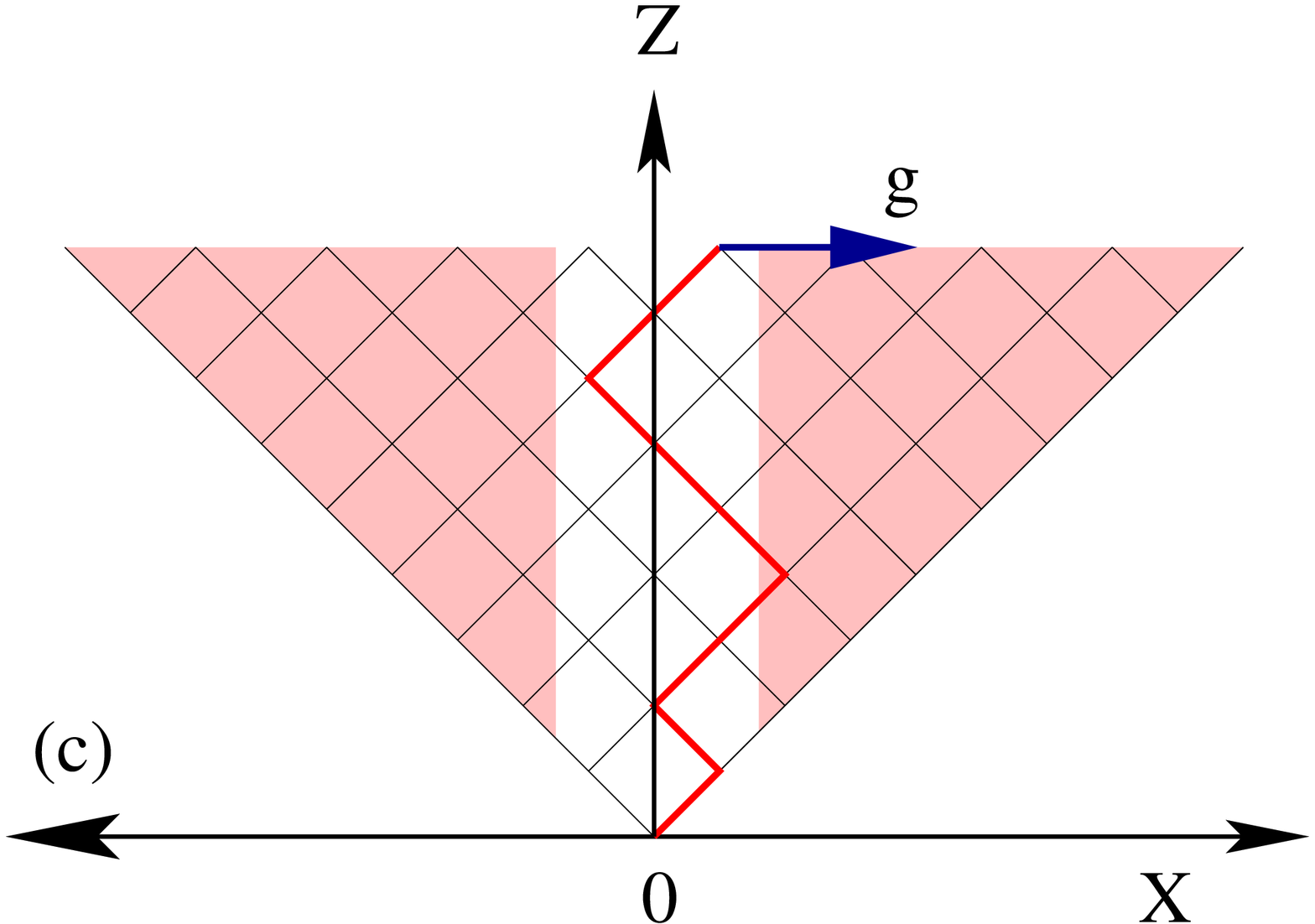}
\includegraphics[width=1.65in]{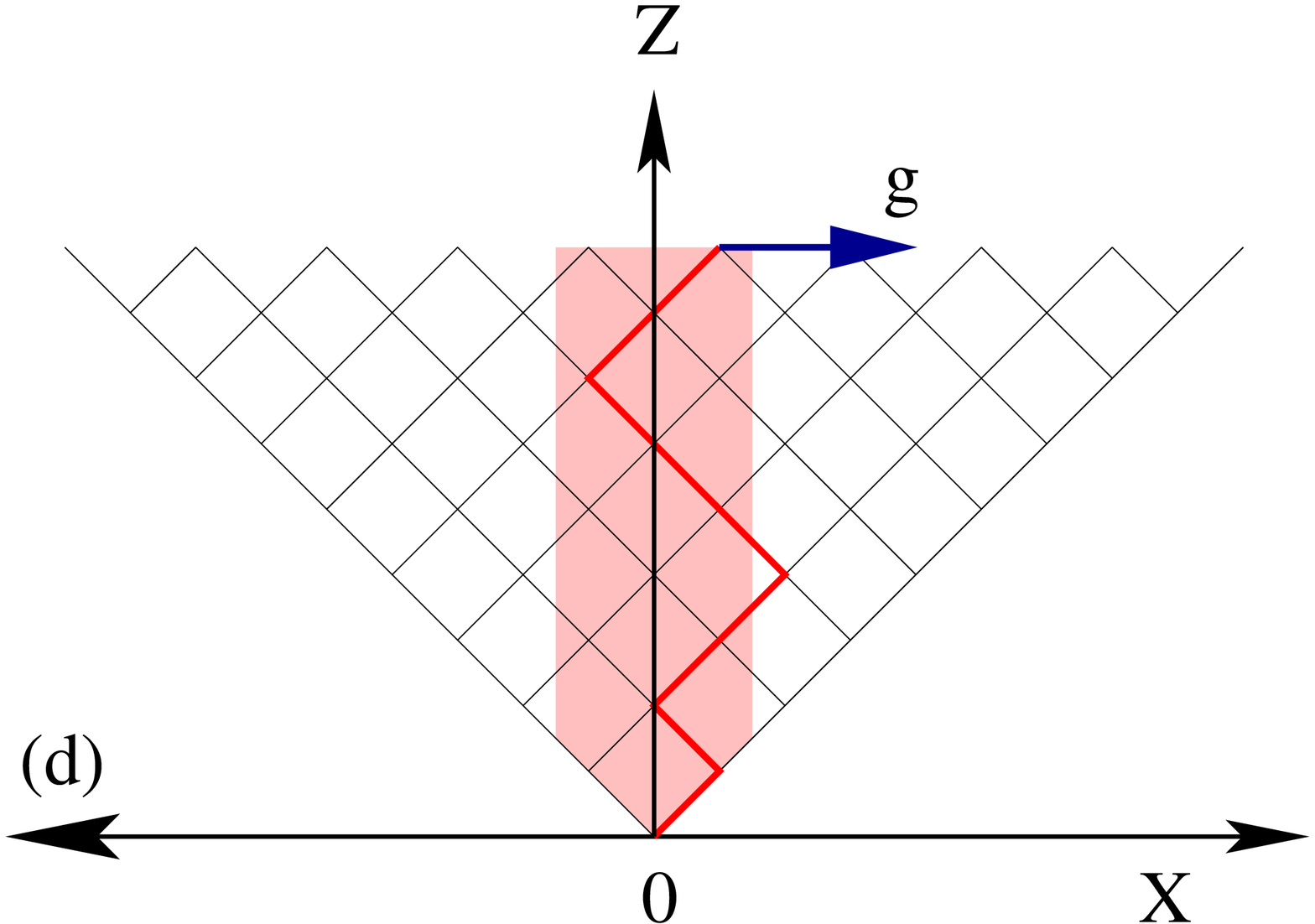}
\caption{Schematic diagram of a directed polymer in a random medium with an attractive wall.  By construction, one direction along the polymer is special. The wandering is in the transverse $d$ directions and therefore it is $d+1$ dimensional model.  Here $d=1$. There is a line at $x=0$ (a wall or an interface) that attracts the polymer. (a) Hard
wall: polymer is not allowed in the region $x<0$. (b) Soft wall: Polymer is allowed in the whole region.  There may be an extra potential $V (>0)$ on one side, say $x<0$. (c) Homo-polymer in random medium. There is a unique (sample independent) ground state for this case. (d) Hetro-polymer in pure medium. The ground state is sample dependent. In (c) and (d),  randomness is in the shaded regions.}
\label{fig:mod}
\end{figure}
}

\newcommand{\indsamp}{\begin{figure}[htbp]
\includegraphics[width=3.2in]{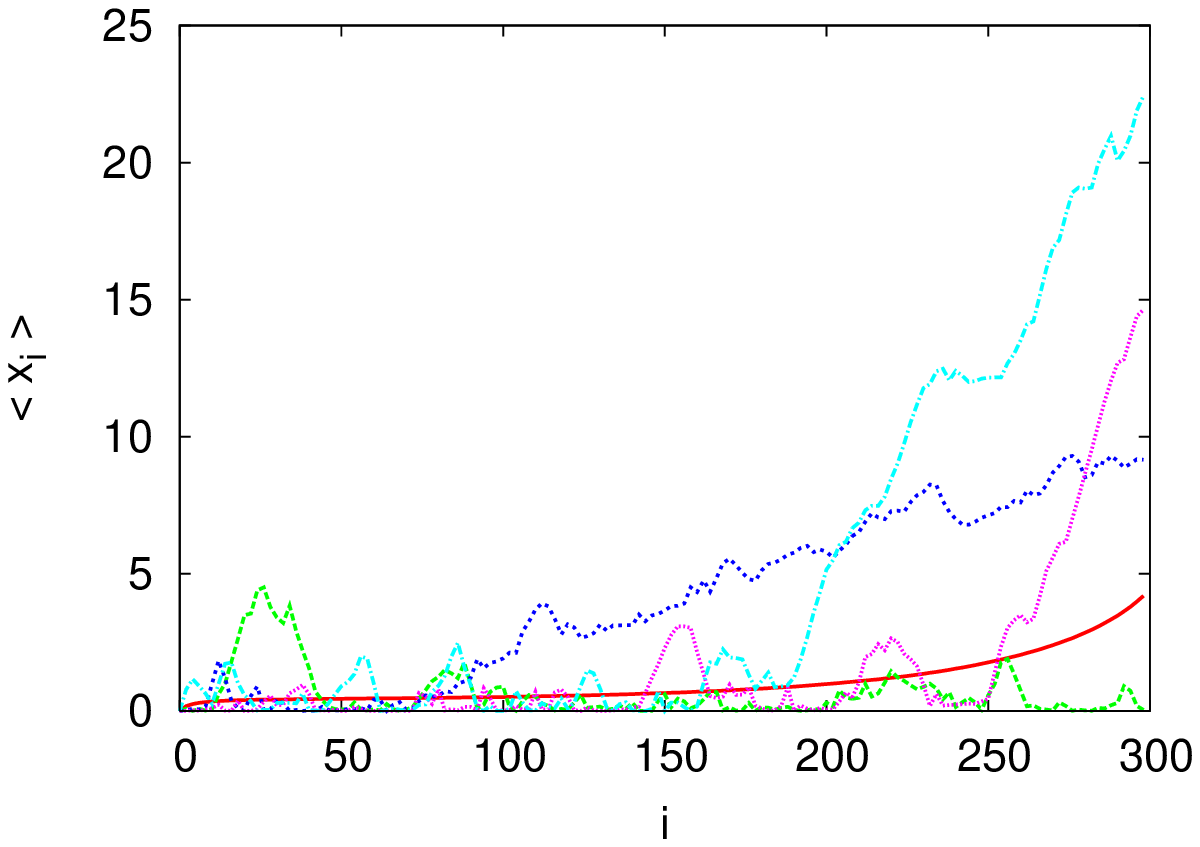}
\caption{ Average separation from the wall, $\langle x_i \rangle$, of $i^{th}$  monomer of chain of length $N=300$ at $\beta \epsilon =3$ $(T = 1/3)$ and disorder strength $\Delta=5/3$  when a stretching force $g=0.5$ is applied at one end. Four different chain configurations are shown by the dotted  and dashed lines. The solid line is when averaging is done over  $90000$ samples.} 
\label{fig:indsamp}
\end{figure}
}

\newcommand{\yforkgc}{\begin{figure}[htbp]
\includegraphics[width=3.2in]{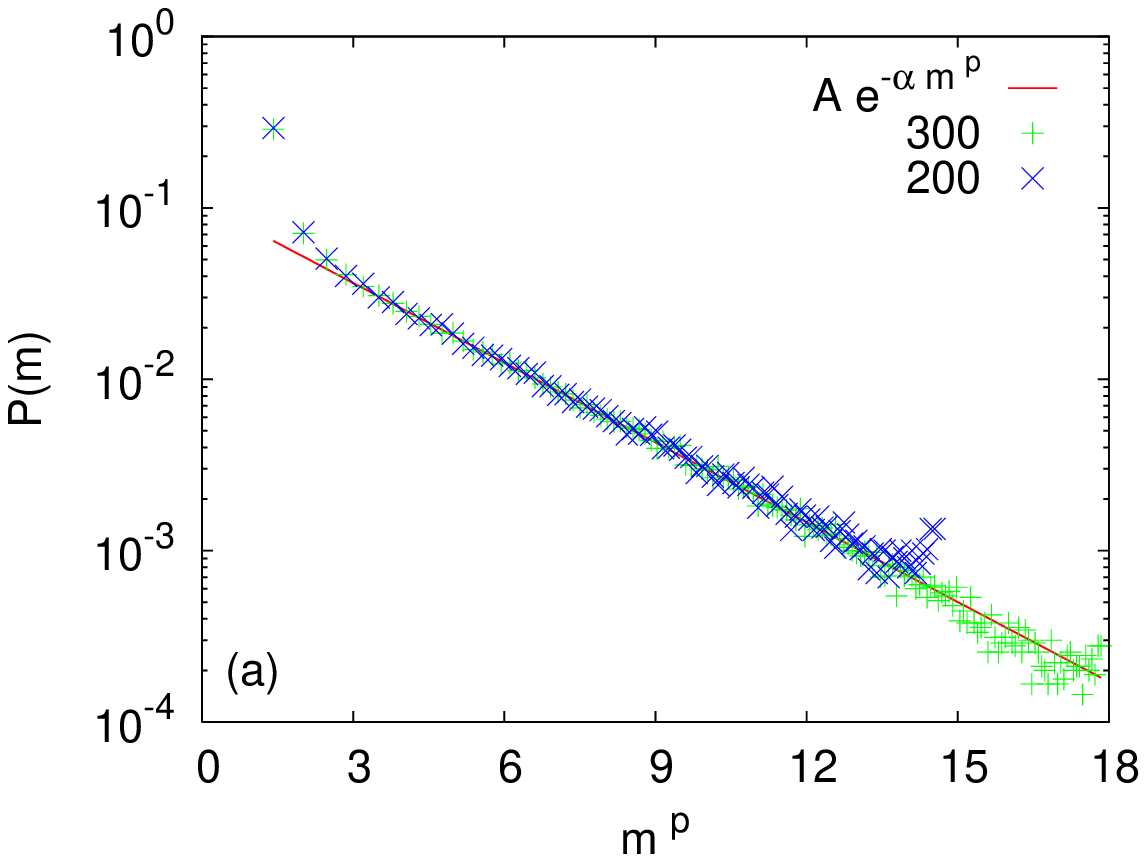}
\includegraphics[width=3.2in]{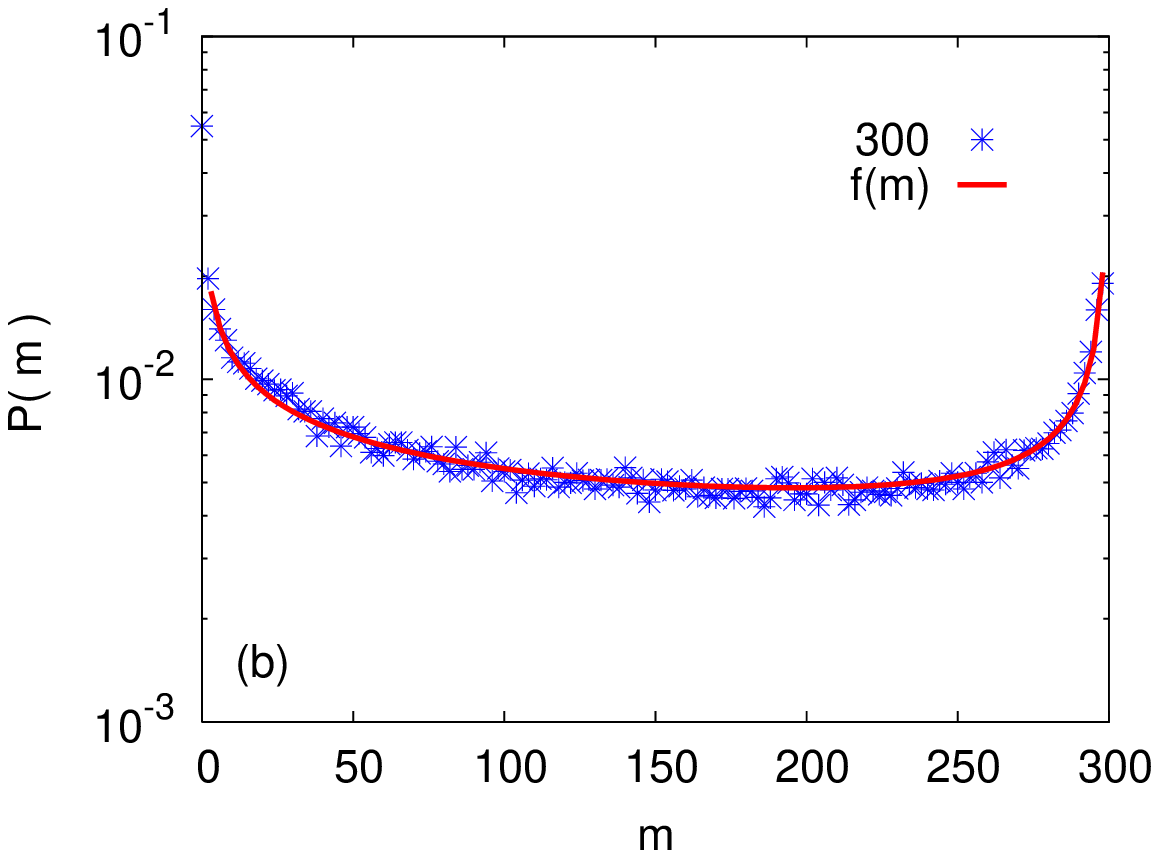}
\caption{ Probability distribution of the unzipped length $m$.  (a)
  Plot of $P(m)$ vs $m^{p}$, at $\beta \epsilon =3$ ($T=0.333)$ and
  disorder strength $\Delta=5/3$, when a stretching force $g=0.5$ is
  applied at the end. The points are for $N =200$ and $300$ and the solid line is the fit using the functional form $f(m) = A e^{-\alpha m^{p}}$. Here $p = 0.51 \pm 0.01$. (b) Plot of $P(m)$ vs  $m$ at  $\beta \epsilon =3$ ($T=1/3$), a  force $g=g_c = 0.82$ is applied. This is the force  on the phase boundary.  The solid curve is the fit using the  functional form $f(m) = c_0/{m^{b_0}} + c_1/{(N-m)^{b_1}}$.To obtain
  the fit, we have excluded $m=0$ and $m=N$. For this plot $c_0=0.026
  \pm 0.001$, $b_0=0.37 \pm 0.0$, $c_1 = 0.028 \pm 0.001$ and $b_1 =
  0.71 \pm 0.02$ are the fitting parameters.  $N=300$ for both the
  figures.  } 
\label{fig:yforkgc}
\end{figure}
}

\newcommand{\bubblegc}{\begin{figure}[htbp]
\includegraphics[width=3.2in]{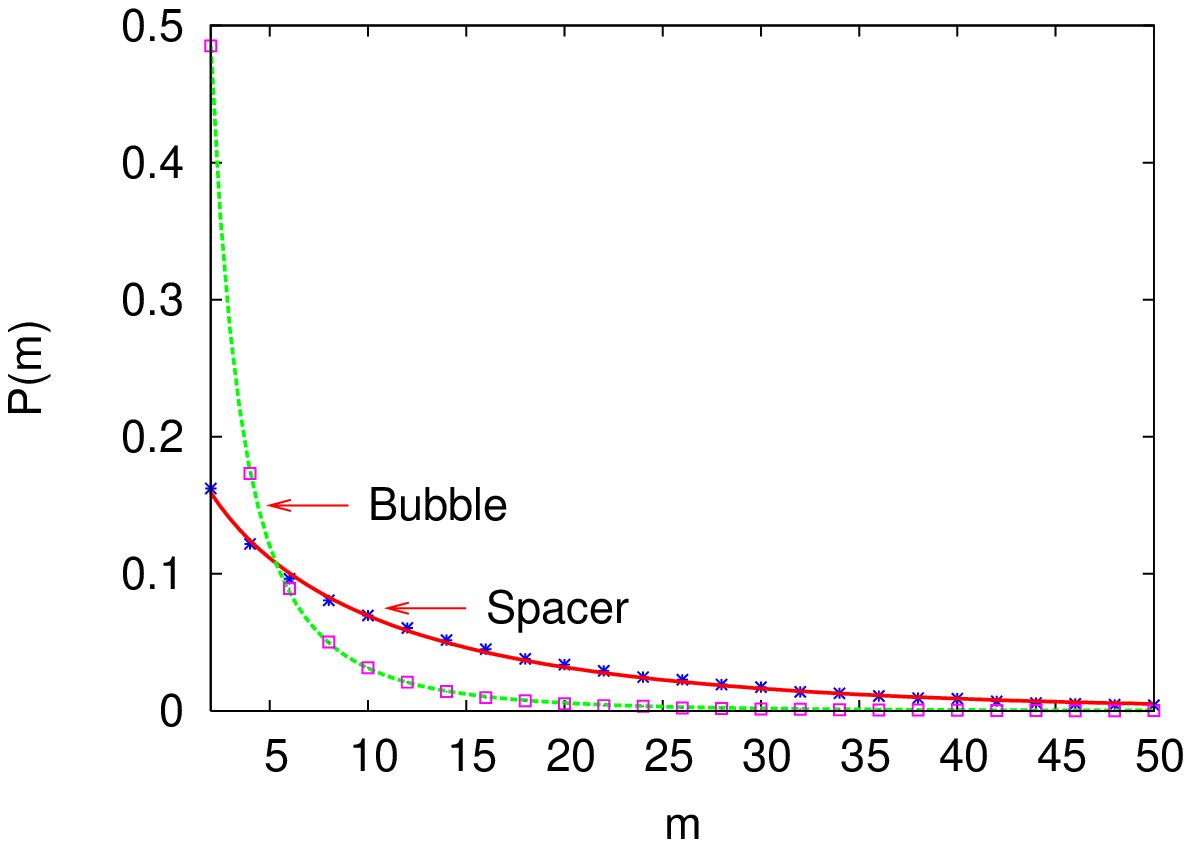}
\caption{Probability distribution of length of the spacer, and the
  length of the first bubble,  
  at critical force $g_c  = 0.82$ and $\beta \epsilon =3$ ($T=1/3$) for a disorder strength
      $\Delta =5/3$. A fitting function of the form $f(m) =
  A e^{-\alpha m ^{c}}$ is used to fit the data. The best fit gives
  the value $c = 0.72 \pm 0.01$ for zipped segment and $c = 0.31 \pm
  0.01$ for the first bubble.}
\label{fig:bubblegc}
\end{figure}
}

\newcommand{\sprand}{\begin{figure}[htbp]
\includegraphics[width=3.2in]{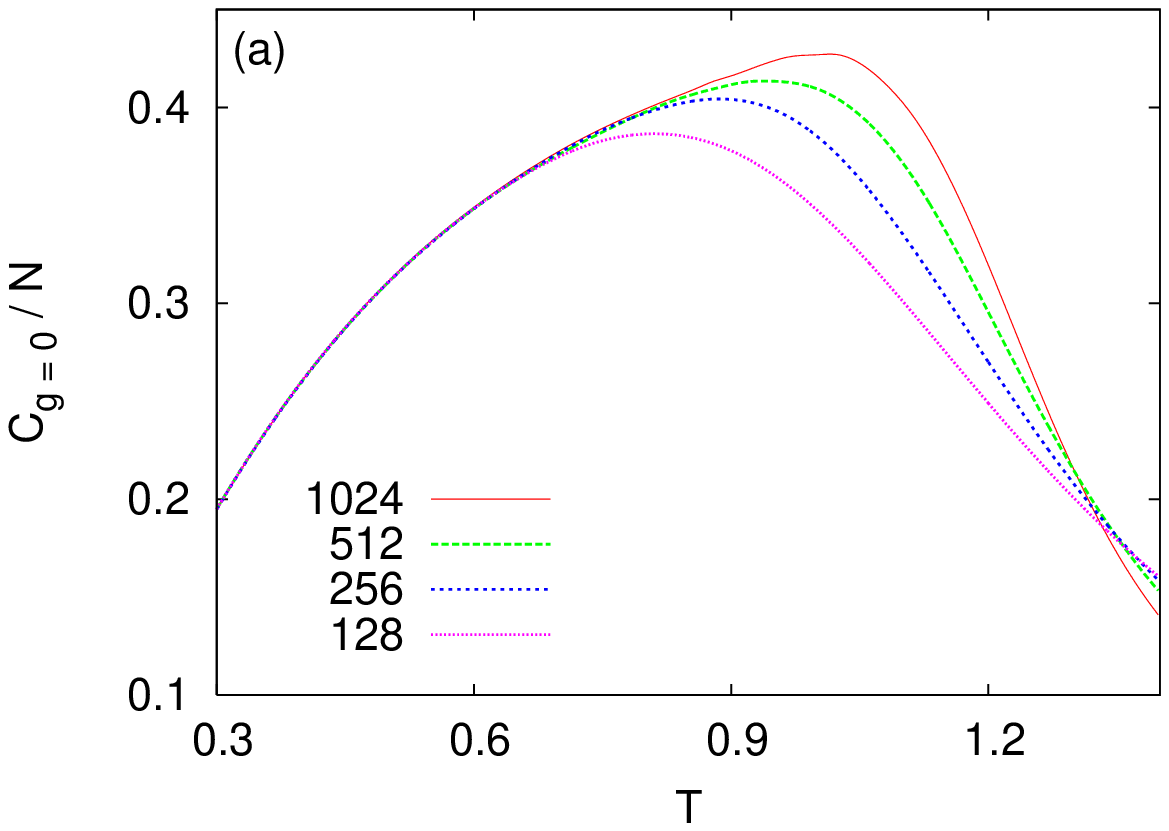}
\includegraphics[width=3.2in]{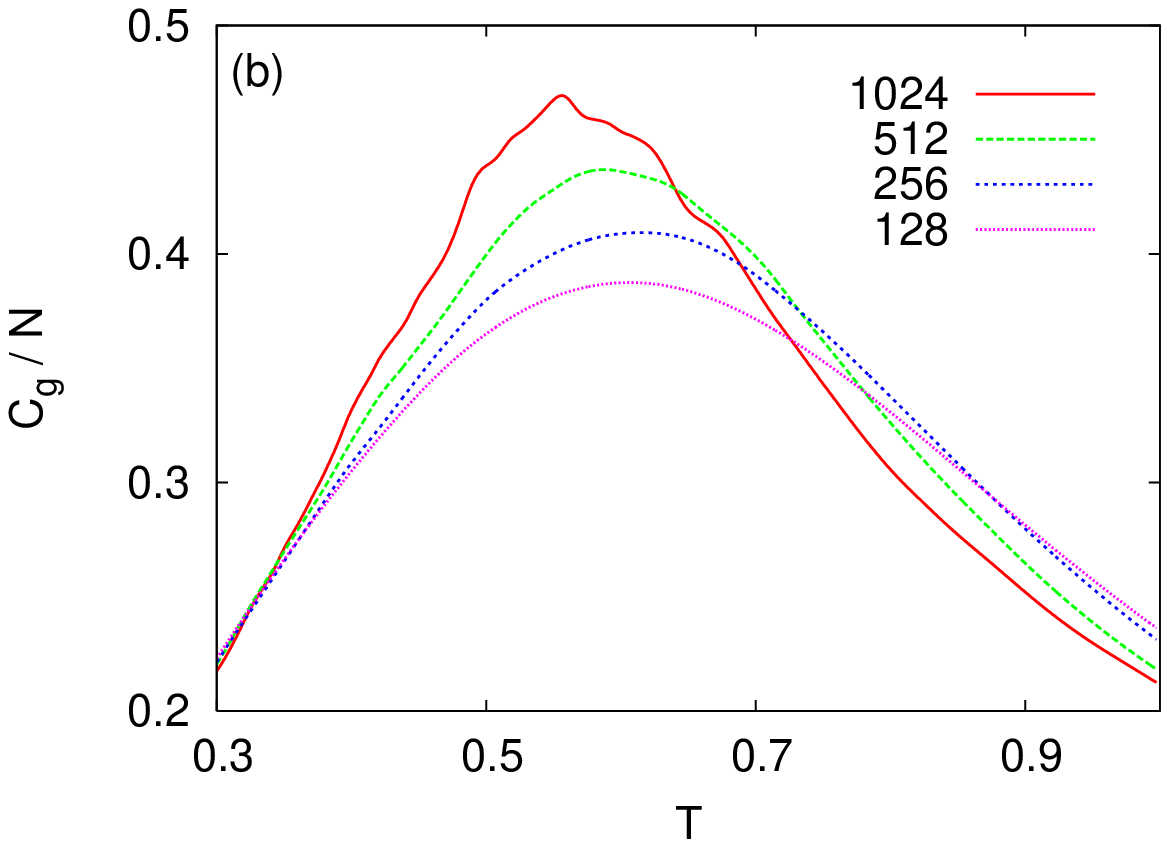}
\caption{Specific heat for the random system at (a)
      $g = 0$ (b) $g =1$. There is 
  a finite size dependence which however does not scale to yield a good data collapse.
  The averaging is done over $10^4$  samples.}
\label{fig:sprand}
\end{figure}
}

\newcommand{\bubzip}{\begin{figure}[htbp]
\includegraphics[width=3.2in]{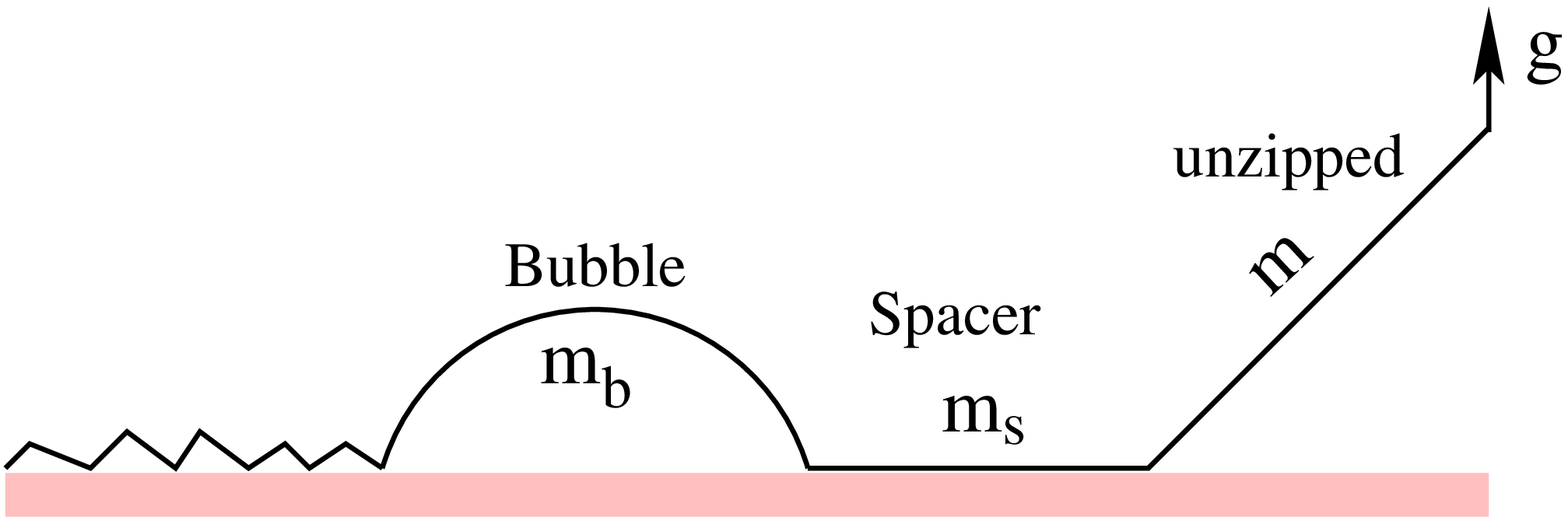}
\caption{Schematic diagram showing the first bubble of length $m_b$ , spacer  of length 
$m_s$ and the unzipped segment of length $m$ when a pulling force $g$ is applied at one 
end of the polymer in the random medium. }
\label{fig:bubzip}
\end{figure}
}

\begin{document}
\bibliographystyle{prsty}      

\title{Unzipping an adsorbed polymer in a dirty or random environment }
\author{Rajeev Kapri and Somendra  M. Bhattacharjee} 
\email{rajeev@iopb.res.in, somen@iopb.res.in}
\affiliation{Institute of Physics,  Bhubaneswar 751005 India.}
\date{\today} 
\begin{abstract}
The phase diagram of unzipping of an adsorbed directed polymer in two dimensions in
a random medium has been determined.  Both the hard wall and the soft wall cases
are considered.  Exact solutions for the pure problem with different affinities
on the two sides are given. The results obtained by the numerical procedure
adopted here are shown to agree with the exact results for the pure case.  The
characteristic exponents for unzipping for the random problem are different from
the pure case. The distribution functions for the unzipped length, first bubble
and the spacer are determined.
\end{abstract}
\maketitle

\section{Introduction}
The study of polymer adsorption on a surface is important because of many
applications such as in lubrication, adhesion, surface protection, coating of
surfaces, wetting, vortex lines, and biology. Of particular interest is the
adsorption-desorption transition of a polymer, similar to the denaturation
transition of a double stranded DNA molecule.  The adsorbed phase, 
dominated by
the attraction with the wall, wins at low temperatures whereas at high
temperatures the desorbed phase prevails.  Although the temperature driven
adsorption-desorption transition dates back to sixties
\cite{rubin:1,privman,yeomans,sumedha,sanjay:sag}, the advent of
micro-manipulation techniques such as optical tweezers, AFM, etc.
\cite{ashkin,essevaz:micro} regenerated interest in it because of the
possibility of exploring single molecules especially under an external force.

The existence of a force induced unzipping transition was established
theoretically for DNA type double stranded molecules in Ref. \cite{smb}.
Many aspects of the unzipping transition have now been elucidated
\cite{sebas,maren:phase,mar:prl,othr,kapri:phase,kumar},  including an
experimental phase diagram \cite{danilo}.  The basic feature of the unzipping
transition is that, at a fixed temperature $T$, the polymers stay in the bound
or localized phase if the magnitude of the pulling force, $g$, is less than the
critical force $g_c(T)$, but they 
can be unbound or unzipped if the pulling force
exceeds the critical force.  Moreover, the transition is described by a set of
characteristic exponents. Since this transition involves a competition between a
binding potential and the stretching by the force, such an unzipping transition
is also possible for an adsorbed polymer \cite{orlandini,pkmishra}.

Apart from thermal fluctuations and force, disorder (or impurities) in
the medium may also induce the adsorption-desorption transition even
below the desorption transition temperature.  This phenomenon is known
as a {\it depinning transition}.  The depinning transition for a
directed polymer has been studied in the past because of its
importance in understanding fluctuations in domain walls in random
magnet, pinning of magnetic flux lines in dirty superconductors having
columnar defects and in particular as a simpler sibling of the more
complex polymer adsorption problem \cite{balent,natt,lipow}. In a
homogeneous medium, in absence of any attraction with the wall, a
polymer behaves like a free chain controlled by entropy.  But with
quenched impurities, there may be one or more ground state
configurations \cite{smb:bkc}.  As a result, a polymer may swell over
a free chain to take advantage of the attractive locations in the
medium.  This swelling effect is pronounced in lower dimensions and
would affect the response of the polymer to any applied force.  More
importantly, because the impurities are quenched (no thermalization
allowed), the behavior of a polymer is sample dependent.  An averaging
over the samples often yields well-behaved thermodynamic quantities
but still sample variations may reveal drastic differences from the
average behavior.  Consequently the unzipping phenomenon in a random
medium (``dirty environment'') is expected to be different from that
in a homogeneous medium because, in the random case, it is not just
pulling off from the wall but also pulling away from local pockets of
favorable configurations.

In this paper our aim is to study the phase diagram of a directed polymer which
is adsorbed on a wall in a random environment and pulled by a force at one end
to unzip it off the wall, keeping the other end fixed.  Though the unzipping
transition in a pure system is first order, randomness may change the order of
the transition.  This change will be reflected in the exponents that describe
the transition. Therefore, in addition to the overall phase diagram we would
also like to determine the exponents in this paper.

\figmodel

The disorder in the problem can be introduced in several ways as shown in Fig.
\ref{fig:mod}. These possibilities are as follows.  (i) The polymer sequence is
homogeneous, and the randomness is introduced on the lattice site such that the
ground state of the polymer (on the wall) remains unaffected as if it is in pure environment.
In this case the binding from the wall is attractive and constant. (ii) The
randomness is introduced in the polymer sequence but the environment is kept
pure. The binding from some portion of the wall, depending on the strength of
randomness, can be attractive or repulsive. The ground state in this case is
sample dependent (iii) The randomness is put at each lattice point including the
wall.  For strong enough disorder some sites of the wall may become repulsive.
In this case also the ground state is sample dependent.  Of these, item (ii) of
heterogeneous sequence is of importance because of its similarity with the DNA
unzipping problem. In this paper we restrict ourselves to case (iii) only,
especially  because of the known behavior of a polymer in a random or dirty
medium.

The thermodynamic or bulk behavior is described by temperature and
one of the two conjugate variables force or distance.  This results
in two different ensembles, namely the fixed force and the fixed
distance ensemble.   For the pure case the ensembles are equivalent
when the polymer is pulled at the end. But this is not always the case. It
is shown recently by Kapri et. al \cite{kapri:phase} in the context of 
DNA unzipping that the ensembles are not equivalent when the
pulling force is applied at any fraction $s$ $(0 < s < 1)$ on the
double stranded DNA. In this respect pulling at the end is a special
case in which results happened to be ensemble independent.  We
restrict ourselves to the force at the end case only and again show
that the pure phase diagram can be obtained exactly from both the
ensembles.  However there are subtle differences and such differences
play a significant role in the random case.  

Apart from exact solutions for the pure case, our approach involves
the use of exact transfer matrix for a polymer on a two dimensional
square lattice.  The choice of two dimensions is partly influenced by
the known results of strong effect of disorder in lower dimensions.
The paper is organized as follows : The model is described in section
II. The quantities of interest are also defined here. In section
III we discuss the adsorption-desorption transition in absence of
randomness.  The exact phase diagram and a comparison with exact
transfer matrix results can be found here.  We also explain the use of
finite size scaling to extract the critical value and the exponents
from the results of extensibility for finite chains.  A general case
of the adsorbing wall separating media of two different types is also
discussed.  Section IV is devoted to the case of a polymer in a random environment.
We determine the phase diagram and the changes in the exponents due to
randomness.  These are based on the results of the transfer matrix for
finite chains and are therefore exact for each sample.  The
statistical error comes only because of averaging over a finite sample
size.  For the data collapse in finite size scaling analysis, the
Bhattacharjee-Seno procedure has been adopted \cite{bh:seno}.

\section{Model} 
We model a polymer by a directed random-walk in a two dimensional
$(d=1+1)$ square lattice directed along the diagonal of the square.
There is an attractive wall on the diagonal of the square ($x=0$)
which tends to suppress wandering because the polymer gains an energy
$-\epsilon$ ($\epsilon>0$) each time it is on the wall \cite{colpin}.
Depending on the ensemble one is working with, either a fixed force, $g$, is
applied at the end monomer ({\it fixed force ensemble}) or the distance,
$x$, of the end monomer from the wall is kept fixed ({\it fixed distance
ensemble}).  The other end is always kept anchored.  Two types of walls
are possible, namely {\it(i)} a hard wall if the polymer can not cross
the wall, and {\it (ii)} a soft wall if the polymer can cross the
wall.  For the random case, there is an additional random energy at
each lattice point.  In a continuum notation, the Hamiltonian may be
written as
\begin{eqnarray}
  \label{eq:20}
  H&=& \frac{d}{2}K \int_0^N dz \left( \frac{\partial
{\bf r}}{\partial z}\right ) ^{^2} + \int_0^N dz V_w({\bf r}(z))
\nonumber\\
&&\qquad + \int_0^N
dz\  \eta({\bf r}(z),z) - 
{\bf g}\cdot \int_0^N \frac{\partial {\bf r}}{\partial z} \ dz
\end{eqnarray}
where ${\bf r}(z)$ is the $d$-dimensional co-ordinate at
monomer index $z$, $V_w({\bf r})$ is the potential due to the wall, $\eta$ is the random energy and ${\bf g}$ is the applied force at the end point $z=N$.  In this paper we consider a discrete version (Fig. \ref{fig:mod}) of this model. 

A more general case can be considered where the polymer experiences
different potentials on the two sides of the wall, as, e.g., may
happen for adsorption at the interface of two immiscible liquids with
different affinities for the polymers. A hard wall case may then
correspond to a solid-liquid interface.  The general case is
considered in the next section and in detail in Appendix A.

\subsection{Pure case}\label{sec:pure}
The partition function for the pure case is obtained by a recursion
relation \beq\label{1} Z_{N+1}( x)= \sum_{j=\pm 1}Z_N(x + j) \left[
1+(e^{\beta\epsilon}-1)\delta_{x,0}\right] \eeq where $Z_N(x)$ is the
canonical partition function of the polymer of length $N$ whose
$N^{\rm th}$ monomer is at a distance $x$ from the wall, the zeroth
monomer being at $x=0$, and $\beta$ is the inverse temperature $1/ T$
in units of the Boltzmann constant $k_{\rm B}=1$.  The temperature
dependence of the partition function has not been shown explicitly.
The initial condition for the above recursion relation of the
partition function is
\begin{equation}\label{eq:7}
Z_0(x)= e^{\beta\epsilon}\delta_{x,0}.
\end{equation}
The non-crossing constraint of the polymer at the  hard-wall is taken care of by reassigning 
\begin{equation}\label{eq:8}
  Z_j(x) = 0, \quad \forall x<0, \qquad {\rm (hard wall)}
\end{equation}
after each step $j$.  

The canonical partition function with a force, $g$, acting at one end
keeping the other end fixed is then calculated by summing over all the
allowed configurations of the walks on the lattice.
\beq\label{2}
{ {\cal{Z}}_{N}(g)=\sum_{x}Z_{N}(x) e^{\beta gx} },
\eeq
where $e^{\beta gx}$ is the Boltzmann weight due to force {\bf $g$}.
The thermodynamic properties in a given ensemble are obtained from the
free energy,
\begin{subequations}
\begin{eqnarray}
F_N(T,x)&=&-T\ln Z_N(x),\qquad {\rm (fixed \ distance)}\label{eq:fnga}\\
{\cal{F}}_N(T,g)&=&-T\ln {\cal{Z}}_N(g),\ \qquad {\rm (fixed\  force).\label{eq:fngb}}
\end{eqnarray}
\end{subequations}

\subsection{Ensembles}
For the fixed distance ensemble, the distance, $x$, of the last (
$N^{\rm th})$ monomer from the wall is kept fixed. The quantity of
interest, the average force required to maintain this distance is
given by
\begin{equation}
  \label{eq:1}
 \langle g \rangle = \frac{\partial F(T,x)}{\partial x} \quad
 (k_{\rm B}=1), 
\end{equation}
and is obtained by the finite difference in free energy $ F(x)$ of
Eq. \ref{eq:fnga} for the lattice problem.

In the fixed force ensemble, the average distance of the last monomer
from the wall, $\langle x \rangle$, where a fixed force, $g$, is
applied is the quantity of interest and can be obtained by
\beq
\langle x \rangle = \frac{\sum_x{ x Z_N(x)e^{\beta gx} }}{\sum_x{Z_N(x) e^{\beta gx}}}.
\eeq

The finite size effects in fixed distance ensemble are weak which
gives the privilege of working with chain of shorter length and still
getting the exact phase boundary. In contrast, the fixed force
ensemble exhibits strong finite size effects and finite size scaling
has to be used to get the phase boundary.

The response of the polymer is characterized by a susceptibility like
quantity to be called {\it isothermal extensibility} defined as
\begin{equation}
  \label{eq:2}
  \chi_T=\frac{\partial \langle x \rangle}{\partial g}.
\end{equation}
This extensibility can be expressed in terms of the  position
fluctuation in the fixed force ensemble as 
\begin{equation}
  \label{eq:4}
  \chi_T = \frac{1}{k_BT} \left ( \langle x^2 \rangle - {\langle x \rangle}^2
  \right ).  
\end{equation}
This relation is useful for numerical computation. We determine the isotherms
$g$ vs $\langle x \rangle$ or $\langle g \rangle$ vs $x$ (depending on the ensemble).

\subsection{Polymer in a random environment}
In the random environment the recursion relation has also the contribution due
to the randomness. The recursion relations in the fixed distance and the fixed
force ensemble respectively read as
\begin{subequations}
\begin{eqnarray}
Z_{N+1}^{\{\alpha\}}(x)
&=& \sum_{j = \pm 1}{ Z_{N}^{\{\alpha\}}(x +j) 
     \left[1+(e^{\beta\epsilon}-1)\delta_{ x,0}\right]}\nonumber\\
&&\qquad \times e^{-\beta\eta(x,N+1)},\label{partdis}
\end{eqnarray}
where $\eta(x,N)$ is the random energy at site ($x,N$), and
\begin{eqnarray}
\label{partfor}
  {\cal{Z}}_{N}^{\{\alpha\}}(g)&=& \sum_{x} e^{\beta g x} \  Z_{N}^{\{\alpha\}}(x).
\end{eqnarray}
\end{subequations}
The superscript $\alpha$ in the above equations denotes a particular
realization.  For every sample, Eq. \ref{eq:7} and \ref{eq:8}  remain valid so
that these recursion relations can be used both for soft-wall  and hard-wall
cases. 

The random energies are independently distributed with no spatial
correlation (white noise), drawn from a uniform deviate of zero mean
$\langle \eta(x,\tau) \rangle = 0$, width $\Delta$, and
\beq\label{randomness}
[\eta(x,\tau)\eta(x^{\prime},\tau^{\prime})]_{\rm dis} 
= \frac{\Delta^2}{12}\  \delta(x-x^{\prime})\,\delta(\tau - \tau^{\prime}).
\eeq
Throughout this paper $[...]_{\rm dis}$  denotes the average
over various realizations and $\langle\ldots\rangle$ denotes the
thermal averaging.

The thermodynamic quantities with quenched randomness come from the
average free energy of the appropriate ensemble, i.e. the average of
${F^{\{\alpha\}}_N(x)}$ for the fixed distance ensemble, and
${{\cal{F}}^{\{\alpha\}}_N(g)}$ for the fixed force ensemble.

With $\epsilon=0$, the randomness leads to a swelling of the polymer
chain as measured by the transverse size
\begin{equation}
  \label{eq:10}
  [<x_N^2>]_{\rm dis} \sim N^{2\nu}, 
\end{equation}
where $\nu=2/3$ in two (1+1) dimensions \cite{smb:bkc}.  For a pure system,
$\nu=1/2$. 

\bubzip

It is apparent that the critical force needed to unzip a polymer has
to be sample dependent. Even in a fixed force ensemble, with $g$ at
its critical value, the whole chain may not be unzipped in all
samples. Important quantities to study are then the unzipped length
$m$ measured from the open end to the first contact (``Y-fork''), the
length $m_s$ of the completely zipped region after that,
to be called a spacer, and then the length $m_b$ of the first bubble.
A bubble is defined as an unzipped region separated by two contacts or
bound monomers.  These are defined in Fig.  \ref{fig:bubzip}. Apart
from the phase diagram obtained by studying the response functions
averaged over samples, the distribution functions of the various
lengths over samples are determined.

\section{Exact solution: Pure case}
\subsection{phase boundary}
The recursion relations, Eqs. \ref{1} and \ref{2}, for the pure case,
can be analyzed exactly.  A more general case where the polymer is
adsorbed at the interface of two different types of media (e.g.
immiscible liquids with different affinity for the polymer) is 
considered \cite{orlandini} and a generating function based derivation 
is given in Appendix A.  Also we show how the phase diagram
can be obtained from a fixed distance ensemble.

The difference in affinity is modeled by a potential $V>0$ for $x<0$.
$V <\infty$ would correspond to two immiscible liquids while the
hardwall $V\to\infty$ corresponds to an interface between a liquid and
an impenetrable solid.  The homogeneous soft wall case corresponding
to the $V\to 0$ limit.  For any $V\neq 0$, there is a temperature
driven desorption transition with $T_c \to \infty$ as $V\to 0$.  The
unzipping transition is given by ($u = e^{-\beta \epsilon}$ and $ v =
e^{-\beta V}$)
\begin{subequations}
\begin{equation}\label{eq:9}
  g_c(T) = T \ln \left [ \frac{ 1 + v (1 - 2 u)}{u( 1  - uv)} - 1 \right ],
\end{equation}
or, equivalently,
\begin{equation}
  \label{eq:3}
  g_c(T)=  2 T \cosh^{-1} \left [ \left \{ 1 - \left ( 1 - \frac{ 2 u ( 1 - v u )}{
  1 + v ( 1 -  2u)} \right )^{2} \right \}^{-\frac{1}{2}} \right ].
\end{equation}
\end{subequations}
To be noted is the small region at intermediate temperatures where a
reentrance can be observed.   The details are given in the Appendix A.

The unzipping on the wrong side, i.e. with $x<0$ where $V>0$, has a
different phase boundary than Eqs. \ref{eq:9} and \ref{eq:3} and is
given by
\begin{equation}
g_c(T)= - 2 T \cosh^{-1} \left [ \frac{1}{v} 
       \left \{ 1 - \left ( 1 - \frac{ 2 u ( 1 - v u )}{1 + v ( 1 -  2u)} \right )^{2} 
       \right \}^{-\frac{1}{2}} \right ].
\end{equation}
The phase diagram is not symmetric around $g=0$, except for $V=0$.
The critical force need not tend to zero at the transition temperature to
unzip on the wrong side.  At $T=0$, the critical force for this model
is $g_c(0)=-2(V+ \epsilon/2)$.  More details of the phase diagram for
$V\neq 0$ will be discussed elsewhere.  

Fig. \ref{fig:phasepure} shows the phase diagram for general $V$,
including the soft-wall and the hard-wall cases. The lines are from
exact results and the symbols from the numerical analysis presented
below. We show in Figs. \ref{fig:fd} and \ref{fig:3}, the final phase
diagrams obtained by the same numerical procedure 
for the random system.
These constitute a part of the major findings of this paper.

\figphasepure
\figfordis
\figthree

\subsection{Units and dimensions}
Throughout the paper we choose dimensionless quantities : Boltzmann
constant $k_B$ is chosen such that $\epsilon/k_B = 1$, and in this
unit, $k_B=1$ with temperature $T$ as dimensionless.  Distances are
measured in units of lattice spacing $a =1$ (along the diagonal).
The quantities like force ($g$), width ($\Delta$) and potential $V$
have dimension of energy and are measured in units of $\epsilon$.  All
the plots shown in this paper use the dimensionless forms where $g
\rightarrow g/\epsilon$, $T$ is $k_BT/\epsilon$, $V \rightarrow
V/\epsilon$, $\Delta \rightarrow \Delta/\epsilon$ and $x \rightarrow
x/a$. In case there is no adsorbtion energy, the relevant parameter is
$\Delta/k_BT$ and the choice of $\epsilon$ does not matter.

\subsection{Isotherms}
For the unzipping transition, $\langle x\rangle$ vs $g$ or $\langle
g\rangle$ vs $x$ isotherms (depending on the ensemble) are of
interest.  The phase diagram can be mapped out from the special
features of these isotherms.  The procedure is explained for the pure
case and the numerical results are compared with the exact results.
This procedure is to be adopted for the random case.

\isotherm

Fig. \ref{fig:isotherm} shows the isotherm for the pure case.  For
finite $N$, these are obtained by iterating the recursion relation
(``transfer matrix'' approach) to calculate the partition function and
then the extension or the force depending on the ensemble used.
Except for a region near $x=0$, the two are identical (within
numerical accuracy).  The difference in the short distance region is
expected because $x$ in fixed distance ensemble has to be an integer
but $<x>$ in the fixed force ensemble need not be \cite{chempot}.

The flat region in the $g$ vs $x$ isotherm signals a coexistence
between a zipped and an unzipped (desorbed) phase.  This is for a
first order transition. The critical value $g_c(T)$ is obtained from 
the intersection of the flat part of the $g$ --- $x$ isotherm with the 
isotherm of the pure phases. For a higher order transition, there may not
be any flat region but some singular feature would persist. In either cases 
singularity on the isotherm gets rounded for finite $N$, and a finite
size scaling analysis can be done to extract the critical value. The advantage 
of using this finite size scaling is that it does not rely on the order of the 
transition.  This method remains valid even if there is no flat part of the 
isotherm (signaling a higher order transition). The exact solution suggests a 
scaling form for extensibility $\chi$,
\beq\label{chipure}
\chi = N^{d_{\chi}} \mathcal{X}\left( (g-g_{c})N^{\phi} \right) 
\eeq
with two characteristic exponents $d_{\chi}$ and $\phi$. As we argue
in Appendix A,  the scaling variable for  the pure problem is
$N(g-g_c)$ so that $\phi=1$.

The scaling variable in Eq. \ref{chipure} can be taken to imply that
the length $m$ of the unzipped region (see Fig. \ref{fig:bubzip}) depends on the force 
as a power law,
\begin{equation}
  \label{eq:12}
  m\sim \mid g-g_c\mid^{-1/\phi}.
\end{equation}
This length $m$ has been used in studying the dynamics of unzipping of DNA \cite{mar:prl}.

If we demand extensivity for $\chi$ for $N\to\infty$, then
$\mathcal{X}(x)\sim |x|^{-(d_{\chi}- 1)/\phi}$.  This implies
\begin{equation}
  \label{eq:5}
  \chi/N \sim |g-g_c|^{-(d_{\chi}- 1)/\phi}.
\end{equation}

A similar scaling form can be written for the specific heat (per
monomer).  At a fixed force, the temperature dependence of the
specific heat is
\begin{equation}
\label{eq:sphtsc}
c_g(T)=N^{\alpha\phi_t} {\mathcal Y}( (T-T_c(g))N^{\phi_t}), 
\end{equation}
which defines the exponent $\phi_t$ in addition to the usual bulk
specific heat exponent $\alpha$.  To recover the bulk behavior,  
${\mathcal Y}(x) \sim \mid x\mid^{-\alpha}$ for large $x$.  
For a first order transition, $\alpha=1$.

The unzipping transition is characterized by the four exponents,
$d_{\chi},\phi, \alpha, \phi_t$. These are known from the exact
solution and we compare with the numerical results.  These are also
the quantities needed for the random system.

\section{Transfer matrix: numerical results for the pure case}
We use the transfer matrix technique in both the ensembles to
calculate the exact partition function numerically from the recursion
relations in Eqs. \ref{1} and \ref{2}.

In Fig. \ref{fig:purecoll}, an example of data collapse for
extensibility $\chi$ for the hard-wall case is shown.  This is at
$\beta\epsilon=15$ $(T=0.067)$ for $N =1000, 1500,$ and $2000$. The critical exponents suggested by collapse \cite{bh:seno} are 
\begin{equation}
d_{\chi}=2, \quad {\rm and} \quad \phi=1.
\end{equation}
consistent with the exact solution in Appendix A. This implies
\begin{equation}
\chi/N \sim |g-g_c|^{-1}, \quad {\rm and}\quad m\sim |g-g_c|^{-1}.
\end{equation}

In the thermodynamic limit, both the ensembles give the same phase
boundary.  By the data collapse method for the extensibility, the
critical force $g_c(T)$ can be determined for a given $T$. Once the
critical force is known, the size ($N$) dependence of the average
distance $\langle x_i \rangle$ of the $i^{\rm th}$ monomer for forces
below and above it can be determined.  We have checked that for $g$
below $g_c$, $\langle x_i \rangle$ decreases very rapidly with $N$
showing that the polymer is adsorbed on the wall whereas for $g$
slightly above $g_c$ the polymer is desorbed from the wall with
$<x_i>\propto i$.

Also from the same calculation the specific heat has been computed.  A
comparison of the specific heat is shown in Fig. \ref{fig:spheat}.   The finite
size scaling variable that one gets from the specific heat is of the form $(T
-T_c(g)) N^{\phi_t}$ with $\phi_t=1$.  Though a good collapse is obtained for
$\alpha=0.93$, but the error bar does not preclude $\alpha=1$ (shown in Fig.
\ref{fig:spheat}b), especially if there are finite corrections.
The development of the $\delta$-function peak ($\alpha=1$) at the
transition point is a vindication of the first order nature of the
unzipping transition.

\purecoll

\figspheat

The critical values obtained from data collapse of the numerical
results are used to construct the phase diagram.  These are shown in
Fig. \ref{fig:phasepure}.  The numerical results (points) are in good
agreement with the analytical results (lines).  
\section {Random case: phase diagram}
To get the phase diagram we need to calculate the critical force,
$g_c(T)$, at different disorder strengths and temperature keeping the
binding to the wall constant. We use the finite size scaling of the
correlation function for that purpose.  For a disorder problem,
depending on the way disorder averaging is done, two different
correlation functions can be defined.
\begin{subequations}
\beq\label{sus}
\chi_{dis} = [ \langle x^2 \rangle - {\langle x \rangle}^2]_{\rm dis}
\eeq
\text{and} 
\beq\label{corr}
\mathcal{C}_{dis} = [\langle x^2 \rangle]_{\rm dis} - [ \langle x
\rangle]^2_{\rm dis}
\eeq
\end{subequations}

For Eq. \ref{sus} the disorder averaging is done at the response
function level of an individual sample whereas for Eq. \ref{corr}, the
disorder averaging is done at the moment level.  The critical
exponents for these two different correlation functions may not be the
same. For the problem in hand, correlation functions scale as
\begin{subequations}
\beq\label{chi}
\chi_{dis}= N^{d_{\chi}} \mathcal{W}\left((g-g_{c})N^{\phi} \right) 
\eeq
\text{and}
\beq\label{cdis}
\mathcal{C}_{dis} = N^{d_{\mathcal C}} \mathcal{W}\left( (g-g_{c})N^{\phi} \right) 
\eeq
\end{subequations}
where $d_{\chi}$ and $d_{\mathcal C}$ are the critical exponents for
$\chi_{dis}$ and $\mathcal{C}_{dis}$ respectively with $\phi$
describing the finite size behavior of the force.  For the random
problem, $\phi$ is expected to be different from the pure case of
Eq. \ref{chipure}.

\figcorr

In the fixed distance ensemble the force, $g^{\{\alpha\}}(x,T)$,
required to maintain a distance $x$ from the wall is computed for each
realization by taking the finite differences in free energy
$ F^{\{\alpha\}}{(x,N) = -k_B T\log{Z^{\{\alpha\}}(x,N)}}$,
which on averaging over samples give the average force, $\langle
g(x,T) \rangle$, required to maintain the distance $x$.
\beq\label{forcenum} \langle g(x,T) \rangle =
[F_{N}^{\{\alpha\}}(T,x+a) - F_{N}^{\{\alpha\}}(T,x)]_{\rm
  dis}/a 
\eeq 
where a is the lattice constant (in our case $a = 1$ ). This has also been used 
to generate the phase diagram or crosscheck the results.

The data collapse for both the correlation functions are shown in fig.
\ref{fig:coll}. The exponents for the hard-wall case, as suggested by the
collapse, are 
\begin{equation}
  \label{eq:13}
 \phi =0.5, \quad  d_{\chi} = 1.5, \quad  {\rm and}\quad  d_{\mathcal C} = 2.0. 
\end{equation}

For both the correlation functions we obtain the same $g_c$ value suggesting
that any of them can be used to calculate the critical force. It is found that the
correlation function at the moment level (Eq. \ref{cdis}) behaves more
smoothly than the response function (Eq. \ref{chi}) and so Eq. \ref{cdis}
can be used to calculate $g_c$, with same accuracy for less number of realizations.

The phase diagram obtained by this determination of $g_c(T)$ for
various $T$ and the dimensionless disorder strength $\Delta$  is shown
in Figs.  \ref{fig:fd} and \ref{fig:3}. For small enough $\Delta$ and
also small $T$, the random energies at the off-wall sites are not
energetically favourable to delocalize the polymer. But still there
are pockets or sites with locally favourable energies.  The pulling
force then has to pull the polymer out of the wall and the local
pockets signalling an increase in the critical force until entropic
effects start dominating or the disorder favors a delocalized state.
Such a thin slice of re-entrance for small $\Delta$ is shown in the
inset in Fig.   \ref{fig:fd}. The full three dimensional surface in
the $g$ -- $T$ -- $\Delta$ space is shown in Fig.  \ref{fig:3} for the
hard and the soft wall cases separately.

\sprand

From the determined values of the exponents, we find that the quenched
averaged extensibility $\chi$, which has to be extensive, should have
the same behavior as the pure case, namely,
\begin{equation}
  \label{eq:6}
 \chi/N \sim |g-g_c|^{-1}, \quad {\rm and}\quad m\sim |g-g_c|^{-2}.
\end{equation}
It is to be noted that for heterogeneous chain this extensibility has
been predicted to be $|g-g_c|^{-2}$, though the finite size exponent
$\phi$ is $1/2$.  In other words, the average behavior of the
extensibility remains pure-like though the size dependence or the
scaling variable is like a heterogeneous chain case.  This is one of
the important conclusions of this paper.  For a soft wall case, as
shown in Fig. \ref{fig:coll}, the exponents are the same as in the
hard wall case.

The zero force specific heat is shown in Fig. \ref{fig:sprand}a.
Unlike the pure case, the discontinuity at the transition temperature
is not clearly visible.  It is interesting to note that all the
specific heat curves for different values of $N$ seem to cross at the
same temperature.  Such crossings generally imply existence of a
certain ``universal'' scale \cite{sgm}.  In the present case this scale
could be due to the proximity of the phase transition point but its
identification remains to be done.  There is a size dependence of the
specific heat near the peak as $N$ increases.  This may be a signal of
a very weak divergence of specific heat or even the formation of a
cusp ($\alpha <0$). Fig. \ref{fig:sprand}b shows the specific heat 
vs $T$ for  $g=1.0$. Here also we see a growth in the peak, but it is 
not as sharp as in the pure case (Fig. \ref{fig:spheat}). No good 
collapse can be obtained. Whether a delta peak is forming as $N \rightarrow
\infty$ is not obvious. Longer chains need to be studied to sort
out these issues.

\section{Sample fluctuations}
\subsection{ Response}
To study the response of the force, the partition function is computed 
for polymer of lengths up to $N=2048$ at a fixed temperature $T$, 
binding to the wall $\epsilon$ and disorder strength $\Delta$.
The average distance of the last monomer, $X$, where a fixed force $g$, 
is applied (fixed force ensemble) is
\beq\label{avdis1}
X = [{\langle x^{\{\alpha\}} \rangle}]_{\rm dis} =
[\sum_{x}{x P^{\{\alpha\}}(x)}]_{\rm dis}
\eeq
where
\beq\label{avdis}
P^{\{\alpha\}}(x) =
\frac{Z_N^{\{\alpha\}}(x)}{\sum_{x}Z_N^{\{\alpha\}}(x)}
\eeq
is the probability of the end monomer of sample $\alpha$ being at a
distance $x$ from the wall.

\figb

Fig. \ref{fig:b} shows the re-scaled force $gN^{1/3}$, versus
re-scaled separation, $ \langle x \rangle/N^{2/3}$ from the wall
(Fig.\ref{fig:b}(a)) and the extensibility (Fig.\ref{fig:b}(b)) for
four different samples of lengths $N=1024$ and $N=2048$ at $\beta
\epsilon =15$ ($T= 0.067$) for disorder strength $\Delta= 2$ when
there is a soft-wall.  The large plateau for small force shows that
the energy gain from the wall favors the adsorption of polymer and
some critical force, which depends on sample, is needed to desorb it.
Above this critical force the polymer does not see the wall but
because of the randomness it can still get trapped to some attractive
sites. The distance from the wall does not increase when the force is
increased by small amount.  The response would then just be the
thermal width of the probability distribution which at low
temperatures is very narrow and independent of $N$.  As a result small
plateaus develop in the isotherm for any specific sample as shown in
the plot \cite{smb:bkc,mezard:steps}.  But because of the possibility
of nearly degenerate but spatially separated states, a fixed force of
right amount may produce a large displacement. Consequently, sharp
jumps appear between plateaus in the isotherms.  The corresponding
extensibility of a sample shows spikes (delta function peaks) where
there are jumps in the force versus distance curves and it is zero
within plateaus. Note the ensemble difference here. Whereas a small
change in $g$ (in the fixed force ensemble) can produce a large change
in $x$, in the fixed distance ensemble, a small change in $x$ would
just explore the neighborhood of the preferred configuration. The
spikes of Fig. \ref{fig:b}(b) remain undetected.  On averaging over
samples, we need to get back extensivity for the extensibility $\chi$.
This is possible because of rare samples with probability $\sim
N^{-2\nu+1}$ having degenerate but spatially separated states
(allowing unzipping at zero-force) \cite{smb:bkc}.

The plot between re-scaled force and re-scaled distance from the wall
and the corresponding extensibility but with $\epsilon=0$, for the
same parameters is shown in Fig. \ref{fig:a}. The average response of
the force is linear at small forces \cite{mezard:steps}.

\figa

\figprobsteps

It is interesting to calculate the probability distribution for
average height of steps, $\Delta x$. Fig. \ref{fig:d}, shows a log --
log plot of $P(\Delta x)$ versus the step height $ {\Delta x}$. Here
we have assumed that the steps are independently distributed.  These
are taken over $10^5$ samples at  $\beta\epsilon = 15$ ($T=0.067$) and
disorder strength $\Delta=2$. The straight line is the best fit which suggests the probability
distribution of the form
\begin{equation}
        P(\Delta x) \sim  {(\Delta x)}^{-p}
\label{pdsteps}
\end{equation}
with $p = 1$.

\figc

In Fig. \ref{fig:c} we have plotted again the re-scaled separation
with re-scaled force for both binding and no binding to the wall but
by keeping the underlying disorder same at a disorder strength
($\Delta=2 $) and $\beta \epsilon = 15$ ($T=0.067$) for $N=1024$. The
steps are for a single sample and the smooth curves are for averages
over $10^4$ samples. For large values of force both the curves merge
showing that the effect of the binding is negligible at large force.
 
\subsection{Pseudo critical force}

\figpseudogc

It is clear from Fig. \ref{fig:b} that in the random environment, the
critical force needed to unzip a polymer depends strongly on sample.
E.g., the displacement for a particular sample would depend on the
deviation $g-g_0$, where $g_0$ is the critical force for that sample.
A random system may have another scale that shows the closeness of
$g_0$ to the average critical force $g_c$.  The scaling behaviour of
this scale can be obtained from the probability distribution of $g_0$
or the finite size scaling of the distribution, $P(g_0) = N^{d_g}
{\cal{F}}( |g_0 - g_c| N^{\phi_g})$ that defines $d_g$ and $\phi_g$.

We take the first jump from the wide plateau around $g=0$ as the
measure of the (pseudo) critical force, $g_0$.  The scaled probability
distribution obtained from the frequencies of $g_0$ over
$10^5$ samples for $N=512$ and  $1024$, is shown in Fig.
\ref{fig:gc}.  The plot suggests $d_g= 0.41$ and $\phi_g = 0.31$,
which also implies a sharply peaked probability distribution.  The
fact that $\phi_g \neq \phi$ (see Fig. \ref{fig:coll}) is suggestive
of different scaling of the two different scales.

\subsection{Bubble, Spacer and Unzipped segments}

In the random medium, the disorder controls the configuration of the polymer. As
already said, the critical force needed to unzip the polymer is sample
dependent. It is interesting to study the distribution of various quantities 
such as bubble length, length of adsorbed or zipped and desorbed or unzipped segments 
with the applied force. 

\indsamp

Fig. \ref{fig:indsamp} shows the average separation from the hard-wall
for each monomer of chain of length $N=300$ at a temperature
$\beta \epsilon = 3$  ($T=1/3 $) and   disorder strength $\Delta= 5/3$ when a stretching
    force $g= 0.5$ is applied at the end.  Four different
chain configurations are shown by the dotted and dashed lines whereas
the curve with solid line is the result of averaging over $90000$ such
chain configurations. Although the applied force here is below the
critical force needed to desorb the polymer yet there are some samples
which are in the desorbed phase.

In this paper we concentrate only on the probability distribution of
(i) the  length  of the desorbed or unzipped segment, (ii) the length
of the adsorbed or zipped segment between  unzipped segment and the
bubble (spacer) and (iii) the length of the first bubble. 
These quantities are defined  schematically in Fig. \ref{fig:bubzip}.

\yforkgc

\bubblegc

The probability distribution (over samples) for the unzipped length
$P(m)$, for the length $m$ of the chain which gets desorbed or
unzipped at the phase boundary is shown in Fig. ~\ref{fig:yforkgc}.
For force below the phase boundary, the probability distribution
decays rapidly with length (Fig. \ref{fig:yforkgc}a) but not so on the
phase boundary (Fig. \ref{fig:yforkgc}b).  For $g<g_c(T)$, the
probability distribution seems to be stretched-exponential, $P(m) \sim
\exp(-a \sqrt{m})$, but the distribution for the phase boundary seems
to admit power laws, especially near the peaks at the two extremes.
Such a two peak structure is expected at a phase
coexistence.

Fig. \ref{fig:bubblegc} shows the probability distributions for the
length of spacer ($m_s$) and the length of the first bubble ($m_b$).
Both the distributions at phase boundary are stretched-exponential
$P(m_s) \sim \exp(-a {m_s}^{c_s})$ with $c_s=0.72 \pm 0.01$ for spacer
length and $P(m_b) \sim \exp(-a {m_b}^{c_b})$ with $c_b=0.31 \pm 0.01$
for the bubble.

\section{Summary}
We obtained the exact phase boundary of unzipping
of a polymer in $1+1$ dimensions from a wall separating two regions of
different affinity for the polymer.  We developed a procedure to
generate the phase diagram from an exact numerical procedure and for
the pure case the numerical results do reproduce the exact results.
The unzipping transition is characterized by a set of exponents
describing the length dependence of the extensibility and the specific
heat.  Numerical results on the exponents also agree with the exact
results, namely $d_{\chi}=2,\phi=1,d_{\mathcal C}=1,\phi_t=1$.

In presence of random impurities distributed independently, taken from
uniform deviate of width $\Delta$, a free polymer wanders more than in
a pure medium.  In addition there is a free energy fluctuation that
grows with the length. It is natural that there is a sample dependence
of the unzipping process but the response under large force also has
inner structure characteristic of the local pockets of favorable
sites. However the steps one sees in the response averages out and one
gets a smooth quenched averaged isotherm.  The phase diagram is
determined both for a hard wall and a soft wall.  The exponents
$d_{\chi}=1.5,\phi=0.5$ are different from the pure.  The behaviour of
the specific heat and finer details of the sample dependence need
further studies. Probability distributions of various relevant quantities are
also obtained. 

\renewcommand{\theequation}{A-\arabic{equation}}
\setcounter{equation}{0}  
\section*{APPENDIX A:  PHASE BOUNDARY FOR PURE CASE } 

In this appendix we give details to calculate the phase boundary for a
pure problem. As stated earlier, we consider a more general case by
assuming that there is a potential $V (V>0)$ at each lattice site on
one side of the wall.  The softwall and hardwall are then the limiting
cases for $V \rightarrow 0$ and $V \rightarrow \infty$ respectively of
the problem. 
The recursion relation satisfied by the canonical partition function in the 
presence of potential is

\begin{eqnarray}
\label{eq1:a}
Z_{N + 1} ( x ) = \left \{ 
                     \begin{array}{l@{,\hfill {\rm for \ }}l}
                         [Z_N ( x  +1 )+Z_N ( x  -1 )] e^{ - \beta V} & x < 0 \vspace{3pt}\\
                         Z_N ( x  +1 )+Z_N ( x  -1 ) & x > 0\vspace{3pt}\\
\label{eq1:b}
                         [Z_N (  +1 )+Z_N ( -1 )] e^{\beta \epsilon} &x = 0 
\end{array} 
\right .\nonumber\\
\end{eqnarray}
with the initial condition $Z_0 ( x ) = e^{\beta \epsilon}\delta_{x,
0}$. The generating function for the partition function, 
\begin{equation}
\label{eq:16}
G( z, x) = \sum_N z^N Z_{N} (x),
\end{equation}
can be taken to be of the form (ansatz)
\begin{equation}
  \label{eq:11}
 G( z, x) = \left \{ \begin{array}{l@{,\hfill {\rm for \ }}l}
                                   \lambda^{x} A& x>0\\
                                  {\lambda^{\prime}}^{-x} A& x<0
                                \end{array} 
                                   \right .
\end{equation}
with $\lambda,\lambda^{\prime}$ and $A$ to be determined.  ($z$
dependence of $\lambda, \lambda^{\prime}$ suppressed).  The root test
of convergence then tells us that the singularity closest to origin 
(on the positive real axis) in the complex $z$-plane determines the
partition function for $N\to\infty$.  For finite $N$, a contour
integration with deformation around the singularities of $G(z,x)$  in
the $z$-plane would yield $Z_N$ from Eqs. \ref{eq:16} and \ref{eq:11}.

Using the ansatz Eq. \ref{eq:11} in Eq. \ref{eq:16}, we get
\begin{subequations}
        \begin{eqnarray}
        \label{eq2:a}
       \frac{\lambda^{\prime }}{z}&=&\displaystyle{  [ \lambda^{\prime -x - 1} +
        \lambda^{\prime -x + 1}]\  e^{- \beta V}, }\quad \text{for} \  x < 0,\\
        \label{eq2:c}
       \frac {\lambda^{x} }{z}&=&\displaystyle{  [ \lambda^{x + 1} + \lambda^{x - 1}]  }, \quad       \text{for} \   x > 0,\\
        \label{eq2:b}
         \frac{A}{z}&=&\displaystyle{ [ (\lambda^{\prime} + \lambda)A
           + \frac{1}{z} ]\  e^{\beta
                \epsilon} },\quad \text{for} \  x = 0,
\end{eqnarray}
\end{subequations}
from which one obtains
\begin{subequations}
\begin{eqnarray}
        \label{lambdaA}
        \lambda &=& \frac{1 - \sqrt{1 - 4 z^2}}{2 z}, \\
        \lambda^{\prime} &=& \frac{1 - \sqrt{1 - 4 z^2 e^{-2 \beta V}}}{2 z  e^{-
        \beta V}},\\
\text{and} \quad 
A &=& \frac{1}{ 1 - (\lambda^{\prime} + \lambda) z e^{ \beta \epsilon}}.
\end{eqnarray}
\end{subequations}
The singularities coming from $\lambda$ and $\lambda^{\prime}$ are
\begin{equation}
  \label{eq:17}
  z_1=\frac{1}{2},\quad {\rm and}\quad
  z^{\prime}_1=\frac{1}{2\exp(-\beta V)}.
\end{equation}
$A$ has the singularity
\begin{equation}
        \label{z2}
        z_2 = \frac{1}{2} \sqrt{ 1 -  \left ( 1 - \frac{2 e^{-\beta \epsilon}( 1
 - e^{-\beta
        V} e^{-\beta \epsilon})}{ 1 + e^{-\beta V}( 1 - 2 e^{-\beta \epsilon})}
        \right )^2}
\end{equation}
which depends on both the adsorption energy and the potential $V$. On
taking the limit $V \rightarrow 0$, the problem reduces to the
softwall case.  On the other hand the problem reduces to a hardwall
case in the limit $V \rightarrow \infty$.  In these limits $z_2$
becomes
\begin{equation}
  \label{eq:14}
z_2 = 
\left \{ \begin{array}{l@{,\hfill {\rm for \ }}l}
                                   \displaystyle{\frac{1}{2} \sqrt{1 - ( 1 - e^{- \beta \epsilon}  )^2 }}& {\rm soft wall}\vspace{3pt}\\
                                  \displaystyle{\frac{1}{2} \sqrt{1 - ( 1 - 2e^{- \beta \epsilon} )^2 }} & {\rm hard wall}
                                \end{array} 
                                   \right .
\end{equation}

The relevant partition function in fixed distance ensemble for large $N$ is
approximated by 
\begin{equation}
  \label{eq:15}
Z_N ( x ) \sim \frac{\lambda^{x}(z_2)}{ z_2 ^{N + 1}},\quad {\rm for}\ x>0,  
\end{equation}
via the contour integration method, with $z_2$ as the closest-to-origin
singularity.   The free energy is then  
\begin{equation}
  \label{eq:18}
 \beta F(x) = N \ln z_2 - x \ln \lambda ( z_2 ), 
\end{equation}
upto $x$-independent additive constant.
The force required to maintain the distance $x$ is given by $g = 2
\partial F / \partial x$.  A factor of 2 is needed as per our
definition of unit length as the diagonal of a unit square of Fig. \ref{fig:mod}.
The phase boundary is then given by 
\begin{subequations}
        \begin{equation}
                \label{eq:11A}
                g_c(T) = -2 T \ln \lambda(z_2), 
\end{equation}
or,
\begin{equation}
g_c(T) = T \ln \left [ \frac{ 1 + e^{-\beta V} (1 - 2 e^{-\beta
\epsilon})}{e^{-\beta \epsilon}( 1  - e^{-\beta V}e^{-\beta \epsilon})} - 1
\right ]
\end{equation}
\end{subequations}
which has been quoted in Eq. (\ref{eq:9}).   On taking the limit $V
\rightarrow 0$ and $V \rightarrow \infty$ one gets respectively the phase
boundary for softwall and hardwall.  
\beq
\label{gsoft} 
g_c(T) = \left \{ 
        \begin{array}{l@{,\hfill {\rm for \ }}l} 
        \displaystyle{ T\ln \left[{2e^{\beta\epsilon} - 1}\right]} & {\rm soft
        wall}\vspace{3pt}\\
        \displaystyle{ T\ln\left[{e^{\beta\epsilon} -1}\right] } & {\rm hard
        wall} \end{array} \right .
\eeq 
The zero force melting takes place at $T_c = \infty$ for softwall and $T_{c} =
\epsilon/\ln{2}$ for the hardwall case.  There is a nonzero $T_c$ for
any $V<\infty$.

In the fixed force ensemble, apart from above mentioned singularities, an
additional force dependent singularity, $z_3(\beta g) = [2\cosh(\beta
g)]^{-1}$, comes from the generating function (grand partition function)
\beq\label{GP}
\mathcal{G}(z,\beta g)=\sum_{N=0}^\infty z^N \sum_{ x}
 Z_N( x) e^{\beta g x}.
\eeq
The phase boundary in the fixed force ensemble  comes from equating two
singularities $z_2 = z_3$.  It is 
\begin{subequations}
\begin{equation}
        g_c(T) = 2 T \cosh^{-1} \left [ \left \{ 1 - \left ( 1 - \frac{ 2 u ( 1 - v u
        )}{ 1 + v ( 1 -  2u)} \right )^{2} \right \}^{-\frac{1}{2}} \right ]
 \end{equation}
with 
\begin{equation}
  \label{eq:19}
u = e^{-\beta \epsilon},\quad {\rm and}\quad  v = e^{-\beta V}.
\end{equation}
On taking the appropriate  limits and with a little simplification, one gets
\begin{equation}
\label{gsoftff}
g_c(T) = 
\left \{ \begin{array}{l@{,\hfill {\rm for \ }}l}
                                   \displaystyle{2 T \tanh^{-1}\left[{1- e^{-\beta\epsilon}}\right]} & {\rm soft wall}\vspace{3pt}\\
                                  \displaystyle{2 T \tanh^{-1}\left[{1- 2e^{-\beta\epsilon}}\right] }     & {\rm hard wall}
                                \end{array} 
                                   \right .
\end{equation}
\end{subequations}
With some algebra one can check that Eqs. \ref{gsoftff} and \ref{gsoft} are identical.

 On pulling the polymer on wrong side ($x < 0$ i.e against the potential V), 
the phase diagram modifies as follows:
In the fixed distance ensemble, the relevant free energy comes from the
partition function $Z_N(x) \sim \lambda^{\prime -x}(z_2)/z_2^{N+1}$ which gives the 
phase boundary 
\begin{equation}
        g_c(T) = 2 T \ln \lambda^{\prime}(z_2)
\end{equation}

In the fixed force ensemble the force dependent singularity, $z_3(\beta g)$, of
the generating function (Eq. \ref{GP}) modifies to $z_3^{\prime}(\beta g) = [ 2v
\cosh (\beta g) ]^{-1}$ which on equating with temperature dependent singularity,
$z_2(\beta \epsilon)$ gives
\begin{equation}
g_c(T)= - 2 T \cosh^{-1} \left [ \frac{1}{v} 
       \left \{ 1 - \left ( 1 - \frac{ 2 u ( 1 - v u )}{1 + v ( 1 -  2u)} \right )^{2} 
       \right \}^{-\frac{1}{2}} \right ].
\end{equation}

The low temperature behaviour of the phase boundary can be understood by a simple analysis. Since the wall is at $x=0$, only the even monomers can be on the wall. For the soft wall case, this means the ground state has a degeneracy of $2^{N}$  for $2N$ steps because both sides of the wall are equally allowed. For small enough $T$, when a length $2m$ is stretched out by the force, the net change in the free energy would involve the loss of the bound state entropy and the gain in the energy due to stretching \cite{maren:phase,mar:prl}. This give $F(m)=m\epsilon -g m + T m \ln 2$, from which the low temperature phase boundary comes out as $g_c(T)= \epsilon + T\ln 2$.

For any $V>0$, the degeneracy is completely lifted as it is energetically not
favourable to be on the negative $x$ side of the wall.  Consequently, there is
no extra  loss of entropy of the bound state at low temperatures. This explains
the difference in the low temperature behavior of the phase boundary betweeen
the softwall and the hard wall case (including $V>0$).

\subsection*{1. Finite size behavior} 
Let us use the ansatz of Eq. \ref{eq:11} in Eq. \ref{GP}, 
for finite $N$, the sum over $x$ goes from $-N$ to $N$. Therefore we
shall have terms of the type $\left [ \lambda(z_2)e^{\beta g} \right ]^N$ where
Eq. \ref{eq:15} has been used. Noting that the critical force is given by Eq.
\ref{eq:11A}, the $N$ dependent term (that vanishes for $N \rightarrow \infty$)
is $e^{N \beta (g-g_c)}$. One can then identify the finite size scaling variable
as $N(g-g_c)$ as quoted in  Eq. \ref{chipure}.

\renewcommand{\theequation}{B-\arabic{equation}}
\setcounter{equation}{0}  
\section*{APPENDIX B:  SPECIFIC HEAT FOR THE PURE PROBLEM  } 
For a given force $g$, the temperature, $T_c(g)$, at which the
transition from the adsorbed
to the  desorbed phase takes place can be calculated by using Eq (\ref{gsoft}). Below
$T_c(g)$ the polymer is adsorbed on the wall and the thermodynamic properties come
from the free energy containing the binding term. The relevant partition
function comes from  $z_2$ of Eq. \ref{eq:14}.
For the hard wall case,  the corresponding free energy is 
$N^{-1} {\cal F}(T,g)= T \ln z_2(T)$.
The specific heat, which is the second derivative of the free energy is
\beq\label{b1}
C(T) = T \frac{d^2 {\cal F}(T,g)}{d T^2} = \frac{\epsilon^{2}
\exp{(\epsilon/{T})}}{2 T^2 \left( \exp{(\epsilon/T)} -1 \right)^2 }
\eeq

When the temperature exceeds $T_c(g)$ the polymer gets desorbed. In this case the
thermodynamics is governed  by the force dependent free energy 
$N^{-1} {\cal F}(T,g) = T \ln z_3$ where $ z_3= 1/[2 \cosh (g/T)]$ is the force dependent
singularity. Therefore the fixed-force specific heat is 
\beq\label{b2}
C_g(T) = T \frac{d^2 {\cal F}(T,g)}{d T^2} = \frac{g^2 {\rm sech}^2{({g}/{T})}}{T^2}
\eeq
There is a discontinuity at $T_c(g)$  but superposed on that there is
a delta function for $g\neq 0$.  A delta function at $T_c(g)$ shows
that the transition is first order 
for $g \ne 0$. In the absence of force ($g=0$),  only the
discontinuity survives and it is then  a classical second order
phase transition.

\end{document}